\documentclass[lettersize,journal]{IEEEtran}
\usepackage{amsmath,amsfonts,amssymb,amsthm}
\usepackage[T1]{fontenc}
\usepackage{times}
\usepackage{algorithmic}
\usepackage{algorithm}
\usepackage{array}
\usepackage{textcomp}
\usepackage{booktabs}
\usepackage{stfloats}
\usepackage{url}
\usepackage{verbatim}
\usepackage{graphicx}
\usepackage{cite}
\usepackage{subfig}
\usepackage{balance}
\usepackage[normalem]{ulem}

\usepackage{acro}

\DeclareAcronym{MIMO}{
  short = MIMO,
  long  = multiple-input multiple-output,
}

\DeclareAcronym{MR}{
  short = MR,
  long  = maximum-ratio,
}

\DeclareAcronym{MMSE}{
  short = MMSE,
  long  = minimum mean-square-error,
}

\DeclareAcronym{LS}{
  short = LS,
  long  = least-squares,
}

\DeclareAcronym{UL}{
  short = UL,
  long  = uplink,
}

\DeclareAcronym{BS}{
  short = BS,
  long  = base station,
}

\DeclareAcronym{SE}{
  short = SE,
  long  = spectral efficiency,
}

\DeclareAcronym{3GPP}{
  short = 3GPP,
  long  = 3rd Generation Partnership Project,
}

\DeclareAcronym{DFT}{
short=DFT,
long=discrete Fourier transform
}

\DeclareAcronym{OBE}
{
short=OBE,
long = optimal bilinear equalizer,
}

\DeclareAcronym{DISC}
{
short=DISC,
long = deterministic instantaneous-SINR combiner,
}

\DeclareAcronym{UaTF}{
  short = UaTF,
  long  = use-and-then-forget,
}

\DeclareAcronym{SINR}{
short = SINR,
long=signal-to-interference-plus-noise ratio
}

\DeclareAcronym{SNR}{
short = SNR,
long=signal-to-noise ratio
}

\DeclareAcronym{PAS}{
short = PAS,
long=power angular spectra
}

\theoremstyle{definition}   

\newtheorem{condition}{Condition}

\newtheorem{lemma}{Lemma}

\newtheorem{remark}{Remark}

\begin{document}

\title{DISC: A Deterministic Instantaneous-SINR Combiner for Uplink Massive MIMO Systems}

\author{Emanuel Wennemo,~\IEEEmembership{Graduate Student Member,~IEEE}, Ki Won Sung, Emil Björnson,~\IEEEmembership{Fellow,~IEEE}

\thanks{E. Wennemo, K.W. Sung, and E. Björnson are with the Department of Communication Systems, KTH Royal Institute of Technology, Sweden. Email: \{wennemo,sungkw,emilbjo\}@kth.se. This work was supported by the IS24-0190 grant from the Swedish Foundation for Strategic Research. A preliminary version of this work has been submitted for review to IEEE ICC 2027 \cite{DISC_D}.}
\thanks{
Claude (Anthropic) assisted in producing the code used for numerical simulations in this paper.}
}



\maketitle

\begin{abstract}
Future Massive MIMO (multiple-input multiple-output) systems will fit more antennas into compact arrays, making minimum mean-square-error (MMSE) combining computationally demanding. Maximum-ratio (MR) combining serves as a cheaper alternative but performs poorly in multi-user scenarios. Existing solutions to this either mimic the MMSE combiner using second-order statistics or, as in the optimal bilinear equalizer (OBE), rely on the loose use-and-then-forget (UaTF) bound. In this paper, we propose DISC, a novel receive combiner that instead maximizes a deterministic approximation of the instantaneous signal-to-interference-plus-noise ratio (SINR), obtained by replacing the random terms in the SINR with deterministic equivalents. This approximation becomes increasingly accurate as the number of antennas grows, and it exploits useful information discarded by the UaTF bound. The computational complexity of DISC matches that of OBE, with the low-complexity variant DISC-D being even cheaper than conventional MR combining. We further prove that the SINR achieved by DISC grows unbounded with the number of antennas, even under pilot contamination, similarly to MMSE combining. Finally, we numerically demonstrate that DISC achieves a sum spectral efficiency close to that of MMSE combining under full pilot sharing, while outperforming other combiners across signal-to-noise ratio, pilot length, and spatial correlation models.
\end{abstract}


\section{Introduction}
\IEEEPARstart{M}{assive} \acs{MIMO} (\acl{MIMO})\acuse{MIMO} systems serve many users simultaneously by equipping a \ac{BS} with a large number of antennas. To achieve high \ac{UL} \ac{SE}, such systems rely on receive combiners to coherently combine the received signals across the antennas \cite{massivemimobook}. One widely considered combiner is the \ac{MR} combiner, which uses the channel estimate as its combining vector \cite{Scalable}. With \ac{MR} combining, the desired signal power is maximized, and the computational burden is limited to channel estimation, making it a natural candidate for single-user scenarios. However, since \ac{MR} does not account for interference, performance degrades as more users are served \cite{Demir_2021}. Even as the number of antennas grows, \ac{MR} combining suffers from the inherent limitation of \textit{pilot contamination}, the phenomenon that occurs when users served in the same time-frequency block share pilot sequences \cite{Marzetta_pilotcon}. In fact, it has been shown that the asymptotic \ac{UL} capacity would remain finite under \ac{MR} combining whenever pilot contamination is present \cite{Marzetta_intro}. The poor performance and fundamental limitations of \ac{MR} combining motivated the use of the \ac{MMSE} combiner, which optimally balances the desired signal strength against interference suppression, thereby maximizing the \ac{SINR}. Under \ac{MMSE} combining, unlimited capacity growth was also shown to be possible as the number of antennas increases, provided there are statistical channel variations over the antenna array that are known and unique to each pilot-sharing user \cite{Unlimited}. Since \ac{MMSE} combining is inherently \ac{SINR}-optimal, it serves as a common benchmark against which other combiners are evaluated \cite{OBE, GMR, Cell_Free_OBE}.

Despite the optimality of the \ac{MMSE} combiner, it is not always practically implementable due to its computational demands \cite{Towards}. One can interpret the \ac{MMSE} combiner as a spatial whitening matrix followed by an \ac{MR} combiner, both of which depend on the instantaneous channel realizations. For each coherence block, computing the whitening matrix requires a matrix inversion whose cost scales cubically with the number of antennas. As future \ac{MIMO} systems are expected to operate in the upper mid-band, enabling many more antennas and users to be simultaneously served than in current systems \cite{FR3}, the high cost of \ac{MMSE} combining will become a major concern and eventually a key implementation bottleneck. Several \ac{MMSE} variants have been proposed to reduce complexity \cite{Towards}, but as long as the whitening matrix relies on instantaneous channel realizations, a per-coherence-block inverse is inevitable. This motivates designing the whitening matrix based on second-order statistics instead, yielding a deterministic matrix that only needs to be recomputed when the second-order statistics change. This then raises the question: How should such a combiner be optimally designed?

Various approaches to designing such a statistics-based combiner have been proposed, each with its own limitations. A natural approach is to take the instantaneous \ac{MMSE} combiner and replace the realization-dependent terms with their expectations \cite{wang2026statisticsapproximationenableddistributedbeamforming}, or with their almost-sure asymptotic limits \cite{GMR}. This heuristic mimics a combiner that is optimal given instantaneous realizations, but does not answer the question of how the deterministic filter should optimally be designed. To answer that question, the \ac{OBE} was proposed, which optimizes the \ac{UaTF} \ac{SINR} over combiners of the shape $\mathbf A_k\mathbf y_k$, where $\mathbf y_k$ is the \ac{LS} estimate of the channel for user $k$ while $\mathbf A_k$ is an arbitrary deterministic matrix \cite{OBE}. The \ac{OBE} and combiners built on the same deterministic bilinear structure have been studied under various fading models and in settings such as cell-free and reconfigurable intelligent surface (RIS)-aided systems \cite{Cell_Free_OBE,syed2024designmultiuserrisaidedstatistical}. By optimizing the widely used \ac{UaTF} bound \cite{massivemimobook}, \ac{OBE} obtains a deterministic \ac{SINR}, which in turn yields the deterministic filter $\mathbf A_k$ that maximizes it. However, since \ac{OBE} is based on the \ac{UaTF} bound on the channel capacity, it also inherits the bound's shortcomings. More specifically, the \ac{UaTF} bound is not necessarily tight, and the self-interference property caused by the ``\textit{forgetting}'' of the channel estimates after the combining vectors are computed leads to the \ac{SINR} saturating for high \ac{SNR} \cite{caire2017ergodicratelowerbounds}. This is consistent with the observation in \cite{OBE} that \ac{OBE} performs close to \ac{MMSE} combining at low \ac{SNR} but diverges significantly from it at higher \ac{SNR}. Furthermore, the \ac{UaTF} bound relies on channel hardening, which cannot be guaranteed in regimes such as those with narrow angular spreads. It also exhibits unexpected behaviors, such as not being invariant under combiner normalization \cite{massivemimobook}. Although the \ac{UaTF} bound is convenient, since it yields a deterministic \ac{SINR}, these drawbacks make it not only an inaccurate performance metric but also a suboptimal design criterion. 

In this work, we avoid the \ac{UaTF} bound and instead optimize a deterministic approximation of the instantaneous \ac{SINR}, obtaining a combiner that matches the low cost of \ac{OBE} while overcoming its shortcomings. The main contributions are summarized as follows:

\begin{enumerate}
    \item We construct a novel deterministic approximation of the instantaneous \ac{SINR} by replacing its random terms with deterministic equivalents, and use it to design \acs{DISC} (\acl{DISC})\acuse{DISC}, the optimal combiner formed by applying a deterministic filter to the \ac{LS} estimate. The resulting deterministic \ac{SINR} closely matches the instantaneous \ac{SINR} at practical numbers of antennas.
    \item We prove that the \ac{SINR} achieved by \ac{DISC} grows unbounded with the number of antennas, even under pilot contamination, given standard linear-independence conditions \cite{Unlimited,OBE} on the users' spatial correlation matrices. Further, we show that \ac{DISC} is equivalent to \ac{MMSE} combining under certain conditions.
    \item We characterize the structural difference between \ac{DISC} and \ac{OBE}, showing how the proposed \ac{DISC} exploits information that is discarded by \ac{OBE} due to the forgetting in the \ac{UaTF} bound.
    \item We show that \ac{DISC} has the same computational complexity as \ac{OBE} and as the prefilter used in \ac{MMSE} channel estimation, and propose a low-complexity variant \ac{DISC}-D, with a lower cost than conventional \ac{MR} combining. We confirm these costs both analytically and in measured runtime.
    \item We numerically demonstrate that \ac{DISC} outperforms competing combiners and tracks \ac{MMSE} combining at high \ac{SNR}, where the alternatives saturate, while \ac{DISC}-D retains much of this performance at a significantly lower computational cost. 
\end{enumerate}
In the preliminary version of this work \cite{DISC_D}, we only considered the low-complexity variant \ac{DISC}-D and did not provide the analytical results presented in this paper.

\textbf{Notation:} We let $\mathbf x$ and $\mathbf X$ denote column vectors and matrices respectively. The expectation operator, transpose, hermitian transpose, vectorization operator, Kronecker product, Hadamard product, and trace operator are denoted by $\mathbb E[\cdot ]$, $(\cdot) ^T$, $(\cdot )^H$, $\mathrm{vec}(\cdot)$, $\otimes $, $\odot$, $\mathrm{tr}(\cdot)$, respectively. The operator $\mathrm{diag}(\cdot)$ maps a vector to a diagonal matrix or the diagonal of a matrix to a vector, depending on the input. Furthermore, the spectral norm and the Frobenius norm are denoted by $\|\cdot \|_2$ and $\|\cdot \|_F$, respectively. Definitions are denoted by $\triangleq$, while $\succeq$ and $\succ$ indicate that a matrix is positive semi-definite and positive definite, respectively. The distribution $\mathbf x\sim \mathcal N_{\mathbb C}(\mathbf 0,\mathbf R)$ corresponds to a circularly symmetric complex Gaussian distribution with spatial correlation matrix $\mathbf R$. The matrix $\mathbf I_M$ denotes the identity matrix with size $M$, while the matrix $\mathbf 0_M$ denotes a $M\times M$ all-zero matrix. Finally, we define $a_n\asymp b_n$ as when $a_n-b_n\to_{n\to \infty} 0$  almost surely, for two infinite sequences of random variables $a_n, b_n$.

\section{System Model}
We consider the \ac{UL} of a single-cell MIMO system with $K$ single-antenna users that transmit simultaneously to a \ac{BS} with $M$ antennas. Traditionally, pilot contamination has mainly been studied in multi-cell systems. In future systems, however, larger \ac{MIMO} configurations will be able to serve more users within each cell while the coherence time shrinks, thereby increasing the pilot demand while reducing the available pilot budget \cite{FR3}. As a result, users may need to share pilots even within a single cell, making pilot contamination relevant in dense single-cell networks. This regime motivates our assumed single-cell model. A special case of this single-cell system is a cell-free/distributed \ac{MIMO} system with a finite number of jointly processed access points, where increasing $M$ corresponds to adding more antennas to each access point.

Under the assumption of block-flat fading, we model the channel to user $k$ in an arbitrary coherence block as
\begin{equation}
    \mathbf h_k \sim \mathcal{N}_\mathbb{C}(\mathbf 0,\mathbf R_k),
\end{equation}
where $\mathbf R_k\in \mathbb C^{M\times M}$ is the spatial correlation matrix of user $k$. As is common in massive MIMO literature, we make the assumption that the spatial correlation matrices for all $K$ users are known at the \ac{BS}. This is justified by the fact that these second-order statistics vary on a much slower timescale than the channel realizations themselves, and techniques for obtaining them have been developed in prior work \cite{Covariance_estimate_neumann,Covariance_estimate_Upadhya}.  

We let each user $k$ transmit the pilot sequence $\boldsymbol \phi_k\in \mathbb C^{\tau\times 1}$, with squared norm $\tau$, and pilot transmit power $\eta_k$. The \ac{BS} then receives the pilot signal $\mathbf Y\in \mathbb C^{M\times \tau}$ expressed as
\begin{equation}
\mathbf Y=\sum_{j=1}^K \sqrt{\eta_j} \mathbf h_j \boldsymbol \phi_j^T+\mathbf N,
\end{equation}
where $\mathbf N\in \mathbb C^{M\times \tau}$ is the receiver noise with each element modeled as $\mathcal N_\mathbb C(0,\sigma^2)$ and statistically independent. The \ac{LS} estimate $\mathbf y_k\in \mathbb C^{M\times 1}$ of the channel vector for a given user $k$ is then obtained by multiplying the received pilot signal by the normalized conjugate of the associated pilot sequence, leading to
\begin{equation}
\mathbf y_k=\sum_{j=1}^K\sqrt{\eta_j}\mathbf h_j  \frac{\boldsymbol\phi_j^T \boldsymbol \phi_k^*}{\sqrt{\tau}}+\frac{\mathbf N \boldsymbol \phi_k^*}{\sqrt \tau},
\label{LS}
\end{equation}
where we note that $\mathbf N \boldsymbol \phi_k^*/\sqrt \tau \sim \mathcal N_\mathbb{C}(\mathbf 0,\sigma^2\mathbf I_M)$. In this work, we limit ourselves to orthogonal pilot design for the purpose of understanding and interpretability, but each pilot might be assigned to multiple users. With such a pilot design, we can reformulate \eqref{LS} as
\begin{equation}
\mathbf y_k=\sum_{j\in \mathcal 
P_k}\sqrt{\eta_j\tau}\mathbf h_j +\mathbf n_k,
\label{LS_orth}
\end{equation}
where $\mathcal P_k$ denotes the set of users that share pilot with user $k$ (user $k$ is included) and $\mathbf n_k$ denotes the noise distributed as $\mathbf n_k\sim \mathcal N_\mathbb C(\mathbf 0,\sigma^2\mathbf I_M)$. Further, note that users who share pilots also have identical noise realizations. With \eqref{LS_orth}, the MMSE channel estimate $\mathbf{\hat h}_k\in \mathbb C^{M\times 1}$ of the channel vector $\mathbf h_k$ is \cite[Eq.~4.5]{Demir_2021}
\begin{equation}
    \mathbf{\hat h}_k=\sqrt{\eta_k\tau}\mathbf R_k\mathbf \Psi_k^{-1}\mathbf y_k,
    \label{MMSE_chest}
\end{equation}
where $\mathbf \Psi_k=\mathbb E[\mathbf y_k \mathbf y_k^H]=\sum_{j\in \mathcal P_k}\eta_j\tau\mathbf R_j+\sigma^2\mathbf I_M.$ The error correlation matrix for user $k$ is then
\begin{equation}
    \mathbf C_k=\mathbb E[(\mathbf h_k-\mathbf{\hat h}_k)(\mathbf h_k-\mathbf{\hat h}_k)^H]=\mathbf R_k-\eta_k\tau\mathbf R_k\mathbf \Psi_k^{-1}\mathbf R_k.
\label{Error_corr}
\end{equation}

By assuming that the channel estimate $\mathbf {\hat h}_k$ for user $k$ is given by the MMSE channel estimate in \eqref{MMSE_chest}, an achievable \ac{UL} \ac{SE} for user $k$ is given as \cite[Theorem~5.1]{Demir_2021}
\begin{equation}
\mathrm{SE}_k=\frac{\tau_c-\tau}{\tau_c}\mathbb E\{\log_2(1+\mathrm{SINR}_k)\}.
\label{Capacity}
\end{equation}
Here, $(\tau_c-\tau)/\tau_c$ is the proportion of the coherence block used for data transmission, referred to as the pre-log factor, and the expectation is with respect to the channel estimates. The \ac{SINR} in \eqref{Capacity} is often referred to as the instantaneous \ac{SINR} and is for user $k$ given by
\begin{equation}
    \mathrm{SINR}_k=\frac{p_k|\mathbf v_k^H\mathbf{\hat h}_k|^2}{\sum_{i\neq k}p_i|\mathbf v_k^H\mathbf{\hat h}_i|^2+\mathbf v_k^H\Big(\mathbf Z+\sigma^2\mathbf I_M\Big)\mathbf v_k}.
    \label{SINR_inst}
\end{equation}
Here, $\mathbf v_k$ is the combining vector for user $k$, $p_k$ and $p_i$ refer to the data transmit powers, and $\mathbf Z$ aggregates the spatial correlations of the unknown parts of all users' channels:
\begin{equation}
    \mathbf Z=\sum_{j=1}^Kp_j\mathbf C_j.
\end{equation}
The widely considered \ac{MR} combiner commonly uses the \ac{MMSE} channel estimate as the combining vector, but it can also use the \ac{LS} estimate in \eqref{LS_orth} \cite{Marzetta_intro} to avoid the computational cost of the \ac{MMSE} channel estimate. These \ac{MR} combiners are denoted by \ac{MR}-\ac{MMSE} and \ac{MR}-\ac{LS}, respectively, shown as 
\begin{equation}
    \mathbf v_k^{\mathrm{MR-MMSE}}=\mathbf{\hat h}_k, \quad \mathbf v_k^{\mathrm{MR-LS}}=\mathbf y_k.
    \label{MR}
\end{equation}
Furthermore, the \ac{SINR}-optimal \ac{MMSE} combiner is defined as
\begin{equation}
    \mathbf v_k^{\mathrm{MMSE}}=p_k\Big(\sum_{i=1}^Kp_i\mathbf{\hat h}_i\mathbf{\hat h}_i^H+\mathbf Z+\sigma^2\mathbf I_M\Big)^{-1}\mathbf {\hat h}_k,
    \label{MMSE_comb}
\end{equation}
 which, inserted into \eqref{SINR_inst}, yields
\begin{equation}
    \mathrm{SINR}_k^{\mathrm{MMSE}}=p_k\mathbf{\hat h}_k^H\Big(\sum_{i\neq k}^Kp_i\mathbf{\hat h}_i\mathbf{\hat h}_i^H+\mathbf Z+\sigma^2\mathbf I_M\Big)^{-1}\mathbf {\hat h}_k.
    \label{SINR_MMSE}
\end{equation}
Although \eqref{SINR_MMSE} depends on the random channel estimates and therefore cannot generally be achieved by deterministic filters, it serves as an achievable upper bound on the instantaneous \ac{SINR}, and an important benchmark against which other combiners can be compared.  

Since \eqref{SINR_inst} depends on random vectors and not just second-order statistics, an alternative bound is commonly used in the literature, referred to as the \ac{UaTF} bound. The core concept behind the \ac{UaTF} bound is that the channel estimates are used for computing the combining vectors but are then forgotten, yielding the following achievable \ac{SE} and deterministic \ac{SINR} \cite[Theorem~4.4]{massivemimobook}:
\begin{equation}
    \mathrm{SE}_k^{\mathrm{UaTF}}=\frac{\tau_c-\tau}{\tau_c}\log_2(1+\mathrm{SINR}_k^{\mathrm{UaTF}}),
    \label{SE}
\end{equation}
{\small
\begin{equation}
    \mathrm{SINR}_k^{\mathrm{UaTF}}=\frac{p_k|\mathbb E[\mathbf v_k^H\mathbf h_k]|^2}{\sum_{i=1}^Kp_i\mathbb E[|\mathbf v_k^H\mathbf h_i|^2]-p_k|\mathbb E[\mathbf v_k^H\mathbf h_k]|^2+\sigma^2\mathbb E[\|\mathbf v_k\|^2]}.
    \label{SINR_UaTF}
\end{equation}}With the \ac{UaTF} bound, it is possible to get an expression of the \ac{SINR} that is fully deterministic and only depends on second-order statistics, which makes it very useful as a design criterion. However, the \ac{UaTF} bound systematically underestimates the achievable \ac{SE}, and when channel hardening is not strong or at high \ac{SNR}, the bounds become looser \cite{caire2017ergodicratelowerbounds,massivemimobook}. We therefore avoid designing the combiner based on the \ac{UaTF} bound in this work.   

In this work, we consider combiners of the form $\mathbf v_k=\mathbf A_k\mathbf y_k$, where $\mathbf y_k$ is the least-squares estimate in \eqref{LS_orth} and $\mathbf A_k\in \mathbb C^{M\times M}$ is a deterministic matrix that does not depend on channel realizations. This family includes several combiners proposed in the prior literature, such as \ac{OBE} \cite{OBE}, \ac{MR}-\ac{LS}, \ac{MR}-\ac{MMSE}, and deterministic variants of \ac{MMSE} combining. Within this family, the goal is to design $\mathbf A_k$ such that the instantaneous \ac{SINR} in \eqref{SINR_inst} is maximized, with \ac{MMSE} combining as the natural benchmark.

\section{Proposed Combiner Design}
In this section, we introduce the almost-sure convergence framework used to derive the proposed \ac{DISC} combiner and present the technical conditions under which it applies, along with a physical interpretation. More specifically, we show that a bounded spectral norm of every user's spatial correlation matrix is sufficient to ensure that the other deterministic matrices arising in the derivation are also spectrally bounded. Under these conditions, we construct a deterministic approximation $\overline{\mathrm{SINR}}_k$ of the instantaneous \ac{SINR} in \eqref{SINR_inst} for combiners of the form $\mathbf v_k=\mathbf A_k\mathbf y_k$, and we maximize this quantity with respect to $\mathbf A_k$ to obtain the \ac{DISC} combiner. We then restrict $\mathbf A_k$ to be diagonal, yielding a low-complexity variant we refer to as \ac{DISC}-D, with its corresponding deterministic \ac{SINR} $\overline{\mathrm{SINR}}^{\mathrm{D}}_k$.

\subsection{Almost-sure convergence framework}
To handle the quadratic terms in \eqref{SINR_inst}, we use the well-known trace lemma, specialized to complex Gaussian vectors \cite[Lemma~B.13]{massivemimobook}. We let $\mathbf x_1,\mathbf x_2,...$ be a sequence of random vectors with $\mathbf x_M\sim \mathcal N_\mathbb C(\mathbf 0_M,\mathbf I_M)$ and let $\mathbf P_1,\mathbf P_2,...$ be a sequence of non-random matrices, with $\mathbf P_M\in \mathbb C^{M\times M}$, that satisfy $\limsup_M\|\mathbf P_M\|_2<\infty$. It then follows that  
\begin{equation}\frac{1}{M}\mathbf x_M^H\mathbf P_M\mathbf x_M\asymp\frac{1}{M}\mathrm{tr}(\mathbf P_M),
\label{Trace_lemma}
\end{equation}
where $\asymp$ denotes that the two sides differ by a quantity that vanishes almost surely as $M\to \infty$.

The normalization by $M$ in \eqref{Trace_lemma} is what makes the difference vanish, so every deterministic equivalent in this work is established for a normalized quantity. However, the individual terms in \eqref{SINR_inst} do not all grow at the same order of $M$, so no single normalization keeps all of them. When assembling the deterministic \ac{SINR}, we therefore drop the normalization factors and retain every term. The order-$M^2$ terms set the asymptotic behavior, but since the order-$M$ terms still contribute significantly at finite $M$, it is beneficial to retain them to achieve an accurate approximation of the instantaneous \ac{SINR}. 

\begin{condition}
    \label{cond:Boundness}
For each user $k$'s spatial correlation matrix $\mathbf R_k\in \mathbb C^{M\times M}$ the following holds 
$$\limsup_M \|\mathbf R_k\|_2<\infty.$$
\end{condition}
This is an established assumption in the massive MIMO literature, which reflects the fact that energy spreads in many spatial dimensions, meaning that no single spatial dimension achieves a gain that grows without bound \cite{Condition_1,Unlimited}. 

Under Condition \ref{cond:Boundness}, and since the spectral norm is sub-multiplicative and sub-additive \cite[Sec.~5.6]{Horn_Johnson_2012}, it follows that $\mathbf \Psi_k=\sum_{j\in\mathcal P_k}\eta_j\tau \mathbf R_j+\sigma^2\mathbf I_M$ and $\mathbf \Psi_k^{-1}$ are spectrally bounded, since $\sigma^2>0$. Similarly, we also have that $\mathbf Z=\sum_{j=1}^K p_j\mathbf C_j$ is spectrally bounded, and we restrict $\mathbf A_k$ to be spectrally bounded as well. Finally, we introduce the cross-correlation matrices
\begin{equation}
    \mathbf{S}_{i,k} \triangleq \mathbb{E}[\mathbf{y}_i \mathbf{y}_k^H] = \begin{cases}
        \mathbf{\Psi}_k & \text{if } i \in \mathcal{P}_k, \\
        \mathbf{0}_M & \text{otherwise},
    \end{cases} \label{S_ik} 
\end{equation}
together with the covariance of the channel estimate $\mathbf \Upsilon_k\triangleq\eta_k\tau\mathbf R_k\mathbf \Psi_k^{-1}\mathbf R_k$, and note that these matrices are all spectrally bounded. 
\subsection{DISC}
\label{DISC_sec}
We would like to find a deterministic matrix $\mathbf A_k$ given by
\begin{equation}
\label{Optimization_problem}
    \arg \max_{\mathbf A_k} \frac{p_k|\mathbf y_k^H\mathbf A_k^H\mathbf{\hat h}_k|^2}{\sum_{i\neq k}p_i|\mathbf{y}_k^H\mathbf A_k^H\mathbf{\hat h}_i|^2+\mathbf y_k^H\mathbf A_k^H\Big(\mathbf Z+\sigma^2\mathbf I_M\Big)\mathbf A_k\mathbf y_k}.
\end{equation}
The maximizer, however, depends on the random realizations through $\mathbf y_k$, $\mathbf{\hat h}_k$, and $\mathbf {\hat h}_i$ and is therefore not deterministic. Instead, we replace each term in \eqref{Optimization_problem} with its deterministic equivalent, drop the normalization factors, and then optimize the resulting deterministic \ac{SINR}. Below, we derive these equivalents term by term.

Since $\mathbf y_k$ is a zero-mean, circularly symmetric, complex Gaussian vector with covariance matrix $\mathbf \Psi_k$, it can be written as $\mathbf y_k=\mathbf \Psi_k^{1/2}\mathbf w_k$, with $\mathbf w_k\sim \mathcal N_{\mathbb C}(\mathbf 0,\mathbf I_M)$ \cite{gallager2013stochastic}. By recalling that $\mathbf{\hat h}_k=\sqrt{\eta_k\tau} \mathbf R_k\mathbf \Psi_k^{-1}\mathbf y_k$ and that $\mathbf \Psi_k$ is Hermitian, we reformulate $\mathbf y_k^H\mathbf A_k^H\mathbf{\hat h}_k$ as \begin{equation}\mathbf y_k^H\mathbf A_k^H\mathbf{\hat h}_k=\sqrt{\eta_k\tau }\mathbf w_k^H\mathbf \Psi_k^{1/2}\mathbf A_k^H\mathbf R_k\mathbf \Psi_k^{-1}\mathbf \Psi_k^{1/2}\mathbf w_k.\end{equation} 
We now apply the trace lemma from \eqref{Trace_lemma} and get
\begin{equation}
\label{Almost_surely_nom_disc}
    \frac{1}{M}\mathbf{y}_k^H\mathbf A_k^H\mathbf{\hat h}_k\asymp \frac{1}{M}\sqrt{\eta_k\tau}\mathrm{tr}(\mathbf \Psi^{1/2}_k \mathbf A_k^H \mathbf R_k\mathbf \Psi_k^{-1}\mathbf \Psi_k^{1/2}).
\end{equation} Restructuring the trace using its cyclic property yields
\begin{equation}
    \mathrm{tr}(\mathbf \Psi^{1/2}_k \mathbf A_k^H \mathbf R_k\mathbf \Psi_k^{-1}\mathbf \Psi_k^{1/2})=\mathrm{tr}(\mathbf A_k^H\mathbf R_k).
\end{equation}
Introducing $a \triangleq \frac{1}{M}\mathbf{y}_k^H\mathbf{A}_k^H\hat{\mathbf{h}}_k$ and $b \triangleq \frac{1}{M}\sqrt{\eta_k\tau}\,\mathrm{tr}(\mathbf{A}_k^H\mathbf{R}_k)$, where $a - b \to 0$ almost surely by \eqref{Almost_surely_nom_disc}, the squared numerator satisfies\footnote{Since $\frac{1}{M}\mathrm{tr}(\mathbf A_k^H\mathbf R_k)$ is bounded under Condition~\ref{cond:Boundness} and $\frac{1}{M}\mathbf{y}_k^H\mathbf A_k^H\mathbf{\hat h}_k\asymp \frac{1}{M}\sqrt{\eta_k\tau}\mathrm{tr}(\mathbf A_k^H\mathbf R_k)$, the quantity $\frac{1}{M}\mathbf{y}_k^H\mathbf A_k^H\mathbf{\hat h}_k$ is almost surely bounded.}
\begin{flalign}
    &\frac{1}{M^2}(|\mathbf{y}_k^H\mathbf{A}_k^H\hat{\mathbf{h}}_k|^2 - \eta_k\tau|\mathrm{tr}(\mathbf{A}_k^H\mathbf{R}_k)|^2)=
     && \nonumber \\ &  |a|^2 - |b|^2
    = \underbrace{(a-b)}_{\to 0}\,\underbrace{\bar a}_{\text{bounded}} + \underbrace{(\bar a - \bar b)}_{\to 0}\,\underbrace{b}_{\text{bounded}},
    \label{Square}
\end{flalign}
where the overline denotes complex conjugation. In each product, an almost surely vanishing factor multiplies a bounded one, so the right-hand side tends to zero and
\begin{equation}
\label{Numerator_disc}
\frac{1}{M^2}|\mathbf y_k^H\mathbf A_k^H\hat{\mathbf h}_k|^2 \asymp \frac{1}{M^2}\eta_k\tau|\mathrm{tr}(\mathbf A_k^H\mathbf R_k)|^2.
\end{equation}

Having found the asymptotic equivalence of the numerator in \eqref{Optimization_problem}, we move on to the denominator, starting with the noise term. Using the representation $\mathbf y_k=\mathbf \Psi_k^{1/2}\mathbf w_k$, we apply \eqref{Trace_lemma} and obtain
\begin{equation}
\label{noise_almost_surely}
\frac{1}{M}\sigma^2\mathbf y_k^H\mathbf A^H_k\mathbf A_k\mathbf y_k\asymp\frac{1}{M}\sigma^2\mathrm{tr}(\mathbf\Psi_k^{1/2}\mathbf A_k^H\mathbf A_k \mathbf \Psi_k^{1/2}).
\end{equation}
By the cyclic property of the trace, this simplifies to $\frac{1}{M}\sigma^2\mathrm{tr}(\mathbf A_k^H\mathbf A_k \mathbf \Psi_k)$. 

Since $\mathbf Z=\sum_{j=1}^Kp_j\mathbf C_j$ is deterministic and spectrally bounded, the same equivalences as for \eqref{noise_almost_surely} apply, with $\mathbf Z$ in place of $\sigma^2\mathbf I_M$. It follows that
\begin{equation}
\label{Z_almost_surely}
\frac{1}{M}\mathbf y_k^H\mathbf A^H_k\mathbf Z\mathbf A_k\mathbf y_k\asymp \frac{1}{M}\mathrm{tr}(\mathbf\Psi_k^{1/2}\mathbf A_k^H\mathbf Z\mathbf A_k \mathbf \Psi_k^{1/2}).
\end{equation}
Applying the cyclic property of the trace, this simplifies to $\frac{1}{M}\mathrm{tr}(\mathbf A_k^H\mathbf Z\mathbf A_k \mathbf \Psi_k)$. 

We have now obtained a deterministic equivalent for all terms in \eqref{Optimization_problem} except for the interference term $\sum_{i\neq k}p_i|\mathbf{y}_k^H\mathbf A_k^H\mathbf{\hat h}_i|^2$, where a more careful treatment is required since pilot sharing leads to correlation between $\mathbf{\hat h}_i$ and $\mathbf y_k$. We apply the linear \ac{MMSE} estimator of $\mathbf y_i$ given $\mathbf y_k$ and decompose $\mathbf y_i$ into the part that is predictable given $\mathbf y_k$ and the corresponding error term $\mathbf e_{i,k}$:
\begin{equation}
    \mathbf y_i=\mathbb E[\mathbf y_i\mathbf y_k^H]\mathbb E[\mathbf y_k\mathbf y_k^H]^{-1}\mathbf y_k+\underbrace{\mathbf y_i-\mathbb E[\mathbf y_i\mathbf y_k^H]\mathbb E[\mathbf y_k\mathbf y_k^H]^{-1}\mathbf y_k}_{\mathbf e_{i,k}}.
\end{equation}
Recalling that $\mathbb E[\mathbf y_k\mathbf y_k^H]=\mathbf \Psi_k$ and $\mathbb E[\mathbf y_i\mathbf y_k^H]=\mathbf S_{i,k}$ as defined earlier, we obtain
\begin{equation}
    \label{LMMSE_y}
    \mathbf y_i=\mathbf S_{i,k}\mathbf \Psi_k^{-1}\mathbf y_k+\underbrace{\mathbf y_i-\mathbf S_{i,k}\mathbf \Psi_k^{-1}\mathbf y_k}_{\mathbf e_{i,k}}.
\end{equation}
Since this is an L\ac{MMSE} estimator, $\mathbf e_{i,k}$ is independent of $\mathbf y_k$, which can also be verified directly by computing $\mathbb E[\mathbf y_k\mathbf e_{i,k}^H]$ and using the fact that the vectors are jointly circularly symmetric Gaussian. The covariance matrix of the error term is
\begin{equation}
\label{Error_LMMSE}
    \mathbb E[\mathbf e_{i,k}\mathbf e_{i,k}^H]=\mathbf \Psi_i - \mathbf S_{i,k}\mathbf \Psi_k^{-1}\mathbf S_{i,k}^H,
\end{equation}
which reduces to $\mathbf 0_M$ when $i\in \mathcal P_k$ and to $\mathbf \Psi_i$ otherwise. This behavior aligns with intuition, since $\mathbf y_i$ is fully predictable given $\mathbf y_k$ when they share a pilot, and $\mathbf y_k$ provides no information about $\mathbf y_i$ when they do not. 

With this, the interference term $\mathbf{y}_k^H\mathbf A_k^H\mathbf{\hat h}_i$ can be written as
\begin{equation}
    \mathbf{y}_k^H\mathbf A_k^H\mathbf{\hat h}_i=\sqrt{\eta_i\tau}\mathbf{y}_k^H\mathbf A_k^H\mathbf R_i\mathbf \Psi_i^{-1}(\mathbf S_{i,k}\mathbf \Psi_k^{-1}\mathbf y_k+\mathbf e_{i,k}),
\end{equation}
where $\mathbf y_{k}$ is independent of $\mathbf e_{i,k}$.
We now expand the squared magnitude $|\mathbf{y}_k^H\mathbf A_k^H\mathbf{\hat h}_i|^2$ as
\begin{flalign}
    & |\mathbf{y}_k^H\mathbf{A}_k^H\hat{\mathbf{h}}_i|^2 = \eta_i \tau \Big( \underbrace{|\mathbf{y}_k^H\mathbf{A}_k^H\mathbf{R}_i\mathbf{\Psi}_i^{-1}\mathbf{S}_{i,k}\mathbf{\Psi}_k^{-1}\mathbf{y}_k|^2}_{\text{Coherent interference}} + && \nonumber \\
    & \underbrace{|\mathbf{y}_k^H\mathbf{A}_k^H\mathbf{R}_i\mathbf{\Psi}_i^{-1}\mathbf{e}_{i,k}|^2}_{\text{Non-coherent interference}} + && \nonumber \\
    & \underbrace{2\Re\big(\mathbf{y}_k^H\mathbf{A}_k^H\mathbf{R}_i\mathbf{\Psi}_i^{-1}\mathbf{S}_{i,k}\mathbf{\Psi}_k^{-1}\mathbf{y}_k\, \mathbf{e}_{i,k}^H\mathbf{\Psi}_i^{-1}\mathbf{R}_i\mathbf{A}_k\mathbf{y}_k\big)}_{\text{Cross-term}} \Big). &&
    \label{Expansion}
\end{flalign}
The cross-term can be split into two cases. When users $i$ and $k$ share a pilot, $\mathbf e_{i,k}\sim \mathcal N_\mathbb C(\mathbf 0,\mathbf 0_M)$, which is a point mass at zero. When they do not share a pilot, $\mathbf S_{i,k}=\mathbf 0_M$ from \eqref{S_ik}. In both cases, the cross-term is identically zero.

For the coherent interference, we take the same steps as for the numerator (see \eqref{Almost_surely_nom_disc}), obtaining
\begin{flalign}
    & \frac{1}{M^2}\eta_i\tau|\mathbf{y}_k^H\mathbf{A}_k^H\mathbf{R}_i\mathbf{\Psi}_i^{-1}\mathbf{S}_{i,k}\mathbf{\Psi}_k^{-1}\mathbf{y}_k|^2\asymp && \nonumber \\
    & \frac{1}{M^2}\eta_i\tau|\mathrm{tr}(\mathbf \Psi_k^{1/2}\mathbf{A}_k^H\mathbf{R}_i\mathbf{\Psi}_i^{-1}\mathbf{S}_{i,k}\mathbf{\Psi}_k^{-1}\mathbf{\Psi}_k^{1/2})|^2, && 
    \label{Coherent_interference_disc}
\end{flalign}
noting that the matrix  $\mathbf \Psi_k^{1/2}\mathbf A_k^H\mathbf R_i\mathbf \Psi_i^{-1}\mathbf S_{i,k}\mathbf \Psi_k^{-1/2}$, arising when we apply $\mathbf y_k=\mathbf \Psi_k^{1/2}\mathbf w_k$, is spectrally bounded under Condition \ref{cond:Boundness}. When users $i$ and $k$ share a pilot, $\mathbf S_{i,k}=\mathbf \Psi_k=\mathbf \Psi_i$, so the trace simplifies to $\mathrm{tr}(\mathbf A_k^H\mathbf R_i)$. When they do not share a pilot, $\mathbf S_{i,k}=\mathbf 0_M$ and the coherent term vanishes. The total coherent interference contribution to the deterministic \ac{SINR} is therefore
\begin{equation}
\label{Final_coherent_int}
    \sum_{i\in \mathcal P_k\setminus \{k\}}p_i\eta_i\tau |\mathrm{tr}(\mathbf A_k^H\mathbf R_i)|^2.
\end{equation} 

The non-coherent term in \eqref{Expansion} requires extra care.  Writing it as $\eta_i\tau|\mathbf y_k^H\mathbf b_{i,k}|^2$, with $\mathbf b_{i,k}\triangleq \mathbf A_k^H\mathbf R_i\mathbf \Psi_i^{-1}\mathbf e_{i,k}$, makes it clear that the quadratic form of $\mathbf y_k$ enters through the rank-one matrix $\mathbf b_{i,k}\mathbf b_{i,k}^H$. The spectral norm of that matrix is $\|\mathbf b_{i,k}\|^2$, which generally grows with $M$ and is not bounded, meaning \eqref{Trace_lemma} does not apply. In fact, the non-coherent interference does not concentrate around its mean when normalized by $M$. If instead normalized by $M^2$, the deterministic equivalent would be $0$, meaning that the non-coherent interference is asymptotically dominated by terms that grow with $M^2$. However, to accurately approximate the instantaneous \ac{SINR} at finite $M$, we retain the term, recall that $\mathbf y_k$ and $\mathbf e_{i,k}$ are independent, and take the expectation with respect to $\mathbf y_k$, conditioned on $\mathbf e_{i,k}$ i.e., 
\begin{equation}\mathbb E[\eta_i\tau |\mathbf{y}_k^H\mathbf b_{i,k}|^2\big|\mathbf e_{i,k}]= \eta_i\tau\mathbf b_{i,k}^H\mathbf \Psi_k\mathbf b_{i,k}.
\end{equation} Substituting back $\mathbf b_{i,k}$ gives a quadratic form in which $\mathbf e_{i,k}$ now enters through a deterministic, spectrally bounded matrix. Applying \eqref{Trace_lemma} again and recalling that $\mathbf e_{i,k}$ is a zero-mean, circularly symmetric, complex Gaussian vector with the covariance matrix $\mathbf \Psi_i-\mathbf S_{i,k}\mathbf \Psi_k^{-1}\mathbf S_{i,k}^H$ then yields
\begin{flalign}
\label{Non_coherent_disc}
    & \frac{1}{M}\eta_i\tau \mathbf e_{i,k}^{H}\mathbf \Psi_i^{-1}\mathbf R_i\mathbf A_k\mathbf \Psi_k\mathbf A_k^H\mathbf R_i\mathbf \Psi_i^{-1}\mathbf e_{i,k}\asymp && \nonumber \\ 
    & \frac{1}{M}\eta_i\tau \mathrm{tr}(\mathbf \Psi_i^{-1}\mathbf R_i\mathbf A_k\mathbf \Psi_k\mathbf A_k^H\mathbf R_i\mathbf \Psi_i^{-1}(\mathbf \Psi_i-\mathbf S_{i,k}\mathbf \Psi_k^{-1}\mathbf S_{i,k}^H)).
\end{flalign}
From \eqref{Error_LMMSE} we have that $\mathbf \Psi_i-\mathbf S_{i,k}\mathbf \Psi_k^{-1}\mathbf S_{i,k}^H=\mathbf 0_M$ when $i\in \mathcal P_k$, and $\mathbf \Psi_i$ otherwise. Applying the cyclic property of the trace, the non-coherent term therefore equals
\begin{equation}
        \eta_i\tau \mathrm{tr}(\mathbf A_k\mathbf \Psi_k\mathbf A_k^H\mathbf R_i\mathbf \Psi_i^{-1}\mathbf R_i)
\end{equation}
when $i\notin \mathcal P_k$ and $0$ otherwise. Recalling that $\mathbf \Upsilon_i=\mathbb E[\mathbf {\hat h}_i\mathbf {\hat h}_i^H]=\eta_i\tau \mathbf R_i\mathbf \Psi_i^{-1}\mathbf R_i$, where $\mathbf {\hat h}_i$ is the MMSE channel estimate of $\mathbf h_i$, the non-pilot-sharing case simplifies to $\mathrm{tr}(\mathbf A_k\mathbf \Psi_k\mathbf A_k^H\mathbf \Upsilon_i)$. With this, the total non-coherent interference contribution is
\begin{equation}
    \label{Final_noncoherent_int}
\sum_{i\notin\mathcal P_k} p_i\mathrm{tr}(\mathbf A_k\mathbf \Psi_k\mathbf A_k^H\mathbf \Upsilon_i).
\end{equation}

The deterministic equivalences obtained above are established at different orders of $M$. The numerator in \eqref{Numerator_disc} and the coherent interference in \eqref{Coherent_interference_disc} are normalized by $M^2$, while the remaining terms are normalized by $M$. As mentioned previously, we drop the normalization factors to retain all terms rather than discard those that still contribute at finite $M$.   

Every term in \eqref{Optimization_problem} now has a deterministic equivalent. To express the deterministic \ac{SINR}, we introduce the vectorized form $\mathbf a_k=\mathrm{vec}(\mathbf A_k)\in \mathbb C^{M^2}$, apply the cyclic property of the trace, and use the identity
\begin{equation}
\label{Kronecker_identity}
  \mathrm{tr}(\mathbf A_k^H\mathbf \Upsilon_i\mathbf A_k\mathbf \Psi_k)=\mathbf a_k^H(\mathbf \Psi_k^T\otimes \mathbf \Upsilon_i)\mathbf a_k.
\end{equation}
By applying the same vectorization to the deterministic equivalents in \eqref{noise_almost_surely} and \eqref{Z_almost_surely}, and writing $|\mathrm {tr}(\mathbf A_k^H\mathbf R_k)|^2$ as 
\begin{equation}|\mathbf a_k^H\mathrm{vec}(\mathbf R_k)|^2=\mathbf a_k^H\mathrm{vec}(\mathbf R_k)\mathrm{vec}(\mathbf R_k)^H\mathbf a_k,
\end{equation}
the deterministic \ac{SINR} takes the form given in \eqref{SINR_disc} at the top of the next page.

\begin{figure*}[!t]
\begin{equation}
\label{SINR_disc}
    \overline{\mathrm{SINR}}_k=\frac{p_k\eta_k\tau \mathbf a_k^H\mathrm{vec}(\mathbf R_k)\mathrm{vec}(\mathbf R_k)^H\mathbf a_k}{\mathbf a_k^H\underbrace{\Big(\mathbf \Psi_k
^T \otimes (\sum_{i\notin\mathcal P_k} p_i\mathbf \Upsilon_i+\mathbf Z+\sigma^2\mathbf I_M) +\sum_{i\in \mathcal P_k\setminus \{k\}}p_i\eta_i\tau \mathrm{vec}(\mathbf R_i)\mathrm{vec}(\mathbf R_i)^H\Big)}_{\mathbf M_k}\mathbf a_k}.
\end{equation}
\end{figure*}
The \ac{SINR} in \eqref{SINR_disc} has the form 
\begin{equation}
\label{Rayleigh_quotient}
    \overline{\mathrm{SINR}}_k=\frac{|\mathbf a_k^H\mathrm{vec}(\mathbf R_k)|^2}{\mathbf a_k^H\mathbf M_k \mathbf a_k},
\end{equation}
where $\mathbf M_k\in \mathbb C^{M^2\times M^2}$ is Hermitian positive definite. The expression in \eqref{Rayleigh_quotient} is a generalized Rayleigh quotient \cite[Lemma~B.10]{massivemimobook}, which is maximized by 
\begin{equation}
\label{a_disc}
    \mathbf a_k^{\mathrm{DISC}}=\arg \max_{\mathbf a_k}\overline{\mathrm{SINR}}_k\propto\mathbf M_k^{-1}\mathrm{vec}(\mathbf R_k),
\end{equation}
with the optimal deterministic \ac{SINR}
\begin{equation}
\label{SINR_disc_opt}
\overline{\mathrm{SINR}}_k^{\mathrm{DISC}}=\max_{\mathbf a_k} \overline{\mathrm{SINR}}_k=\eta_k\tau p_k\mathrm{vec}(\mathbf R_k)^H\mathbf M_k^{-1}\mathrm{vec}(\mathbf R_k),
\end{equation}
and thus we have obtained the \ac{DISC} and the maximum achievable deterministic \ac{SINR}.

Although the deterministic equivalents are established as $M\to \infty$, the filter in \eqref{a_disc} is deterministic and can be applied at any finite $M$. The proportionality in \eqref{a_disc} reflects the scale invariance of the instantaneous \ac{SINR} in \eqref{SINR_inst}. Any non-zero scaling of $\mathbf v_k$ leaves the \ac{SINR} unchanged, and $\overline{\mathrm{SINR}}_k$ inherits this property.

\begin{remark} \label{Robust_DISC_remark}
Instead of the shape $\mathbf A_k\mathbf y_k$, the combiner can be designed as $\mathbf{\tilde{A}}_k\mathbf{\hat h}_k$, acting on the \ac{MMSE} channel estimate $\mathbf {\hat h}_k=\sqrt{\eta_k\tau}\mathbf R_k\mathbf \Psi_k^{-1}\mathbf y_k$ instead of the least-squares estimate $\mathbf y_k$. When the spatial correlation matrix $\mathbf R_k$ is invertible, the two shapes are equivalent, since $\mathbf R_k\mathbf \Psi_k^{-1}$ is then a bijective transform. In contrast, when $\mathbf R_k$ is numerically rank-deficient (i.e., when its eigenvalue spectrum decays rapidly), they differ, since the prefilter $\mathbf R_k\mathbf \Psi_k^{-1}$ confines the combiner to the limited numerical range of $\mathbf R_k$. This confinement implies that the combiner cannot suppress interference in directions outside the numerical range of $\mathbf R_k $, which can reduce performance. At very high \ac{SNR}, however, this can provide numerical stability, since the combiner is not forced to create deep nulls in ill-conditioned directions outside the dominant desired signal subspace, as the prefilter has already suppressed the input there. Due to its greater generality, we consider only the shape $\mathbf A_k\mathbf y_k$ in this work, but note that the \ac{MMSE}-estimate-based shape is also a possible implementation.
\end{remark}
Computing $\mathbf a_k^{\mathrm{DISC}}$ requires inverting $\mathbf M_k\in \mathbb C^{M^2\times M^2}$, and although this cost can be reduced by applying the Woodbury identity (as discussed in Appendix A), it may not always be feasible. We therefore propose a low-complexity variant in the next subsection.
\subsection{DISC-D}
\label{sec:DISC_D}
A low-complexity variant of \ac{DISC} can be obtained by restricting $\mathbf A_k$ to be a diagonal matrix, denoted $\mathbf D_k$. Since the combiner shape $\mathbf D_k\mathbf y_k$ is a special case of $\mathbf A_k\mathbf y_k$, the deterministic equivalents derived in the previous section apply directly with $\mathbf A_k$ replaced by $\mathbf D_k$. By defining $\mathbf d_k=\mathrm{diag}(\mathbf D_k)$ the identity in \eqref{Kronecker_identity} reduces to
\begin{equation}
    \mathrm{tr}(\mathbf D_k^H\mathbf \Upsilon_i\mathbf D_k\mathbf \Psi_k)=\mathbf d_k^H(\mathbf \Psi_k^T\odot \mathbf \Upsilon_i)\mathbf d_k.
\end{equation}
Furthermore, with the diagonal constraint, $|\mathrm{tr}(\mathbf D_k^H\mathbf R_k)|^2$ becomes
\begin{equation}
    |\mathrm{tr}(\mathbf D_k^H\mathbf R_k)|^2=\mathbf d_k^H\mathrm{diag}(\mathbf R_k)\mathrm{diag}(\mathbf R_k)^H\mathbf d_k.
\end{equation}
Applying these identities to the remaining terms in \eqref{Optimization_problem} yields the deterministic \ac{SINR} given in \eqref{SINR_disc_d} at the top of this page.
\begin{figure*}[!t]
\begin{equation}
\label{SINR_disc_d}
    \overline{\mathrm{SINR}}^\mathrm{D}_k=\frac{p_k\eta_k\tau \mathbf d_k^H\mathrm{diag}(\mathbf R_k)\mathrm{diag}(\mathbf R_k)^H\mathbf d_k}{\mathbf d_k^H\underbrace{\Big(\mathbf \Psi_k
^T \odot (\sum_{i\notin\mathcal P_k} p_i\mathbf \Upsilon_i+\mathbf Z+\sigma^2\mathbf I_M) +\sum_{i\in \mathcal P_k\setminus \{k\}}p_i\eta_i\tau \mathrm{diag}(\mathbf R_i)\mathrm{diag}(\mathbf R_i)^H\Big)}_{\mathbf M_{k}^D}\mathbf d_k}.
\end{equation}
\hrulefill
\end{figure*}
This is again a generalized Rayleigh quotient, now maximized by \begin{equation}
\label{disc_d_combiner}
    \mathbf d_k^{\mathrm{DISC-D}}\propto (\mathbf M_{k}^D)^{-1}\mathrm{diag}(\mathbf R_k),
\end{equation}
yielding the optimal deterministic \ac{SINR}
\begin{equation}
\max_{\mathbf d_k} \overline{\mathrm{SINR}}_k^\mathrm{D}=\eta_k\tau p_k\mathrm{diag}(\mathbf R_k)^H(\mathbf M_{k}^D)^{-1}\mathrm{diag}(\mathbf R_k),
\end{equation}
denoted by $\overline{\mathrm{SINR}}_k^{\mathrm{DISC-D}}$.

Since $\mathbf M_k^D\in \mathbb C^{M\times M}$ and $\mathrm{diag}(\mathbf R_k)\in \mathbb R^M$, restricting \ac{DISC} to the shape $\mathbf D_k\mathbf y_k$ already reduces the computational cost significantly. Under a uniform horizontal linear array structure, further reductions are available in \ac{DISC}-D. With antenna spacing $d$ in wavelengths, element $(l,m)$ of the spatial correlation matrix is given by \cite{Correlated_fading}
\begin{equation}
[\mathbf{R}_k]_{l,m}=\beta_k\int_{-\infty}^\infty e^{j2\pi (l-m)d  \sin(\overline\varphi_k)}f(\overline \varphi_k)d \overline \varphi_k,
\label{Spat_cor}
\end{equation}
where the total average gain of user $k$ corresponds to $\beta_k$ while $f(\overline{\varphi}_k)$ represents the probability density function of the angle of arrival. Since $\mathbf R_k$ in \eqref{Spat_cor} is Toeplitz, it is asymptotically equivalent to a circulant matrix as $M\to \infty$, according to Szegő's theorem \cite{Szego}. As circulant matrices are diagonalized by the \ac{DFT} matrix $\mathbf F\in \mathbb C^{M\times M}$, we have $\mathbf R_k\approx\mathbf F\mathbf \Lambda_k\mathbf F^H$ with equality as $M\to \infty$, where $\mathbf \Lambda_k$ is a diagonal matrix containing the \ac{DFT}-domain eigenvalues of $\mathbf R_k$. We therefore transform all spatial correlation matrices to the \ac{DFT} domain, also referred to as the beamspace, approximate them as diagonal, and compute the combiner according to \eqref{disc_d_combiner}. For finite $M$, this diagonalization is not exact, so the resulting combiner is an approximation of \eqref{a_disc} that trades performance for a reduction in complexity. Since $\mathbf \Psi_k$, $\mathbf Z$, $\mathbf \Upsilon_i$ for all $i$ are also sums or products of the spatial correlation matrices, they inherit the same approximate diagonalization. With this approximation, $\mathbf M_k^D$ takes a diagonal-plus-low-rank structure, where each pilot-sharing interferer contributes a rank-1 term in $\sum_{i\in \mathcal P_k\setminus \{k\}} p_i\eta_i\tau \mathrm{diag}(\mathbf R_i)\mathrm{diag}(\mathbf R_i)^H$. Due to this structure, it can be inverted efficiently using the Woodbury identity. Applying the combiner to a channel realization is also cheaper than for the general \ac{DISC}, since $\mathbf D_k\mathbf y_k$ reduces to a pointwise multiplication in the \ac{DFT} domain. A more detailed complexity analysis of \ac{DISC}-D is provided in Appendix A. 
\section{Analysis and Comparisons}
\label{sec:Analytical}
In this section, we provide analytical comparisons of \ac{DISC} against the most prominent prior combiners, namely \ac{MMSE} combining, \ac{OBE}, and the \ac{MR} combiners. We first analyze the computational cost, showing that \ac{DISC} matches the cost of \ac{OBE} while avoiding the per-realization inverse of \ac{MMSE} combining, with \ac{DISC}-D achieving an even lower cost. Next, we show that \ac{DISC} is equivalent to \ac{MMSE} combining under certain conditions and characterize how it differs from \ac{OBE}. We then show that the \ac{SINR} achieved by \ac{DISC} grows unbounded over antennas as $M\to \infty$, under standard linear-independence conditions on the spatial correlation matrices, even under pilot sharing. Finally, we establish that the deterministic \ac{SINR} that \ac{DISC} optimizes converges to the instantaneous \ac{SINR} as the number of antennas grows, provided that pilot contamination is present.
\subsection{Computational analysis}
The practical value of a combiner is set not only by the \ac{SINR} it achieves but also by whether it can be computed at all for a given combination of computing platform and number of antennas. In massive \ac{MIMO}, the number of antennas $M$ is large, so costs that scale steeply with $M$ determine feasibility. \ac{MMSE} combining requires inverting an $M\times M$ matrix for every channel realization, at a cost that grows cubically with $M$, which can become too computationally demanding \cite{Towards}.

The computational cost of a combiner is best understood on two time scales. Channel realizations change every coherence block, so any computation that depends on the instantaneous realizations is repeated on a per-realization basis. The spatial correlation matrices are fixed across many coherence blocks, so statistics-based computations, such as the design of \ac{DISC}, are performed once and then treated as fixed. The number of blocks over which the statistics remain fixed depends on the configuration and can range from 40 coherence blocks \cite{40} to 25000 \cite{25000}. Nevertheless, the statistics-based computation should also remain as inexpensive as possible so that it does not become a bottleneck.

In Appendix A, we show how we can reduce the design cost of \ac{DISC} and \ac{DISC}-D, over all users, to $\mathcal{O}(M^3K)$ and $\mathcal{O}(MK\log M)$ respectively, under the assumption that $M$ dominates the cost. These match the design costs of the statistics-based combiners \ac{OBE} and \ac{OBE}-D \cite{OBE}. On the per-realization time scale, applying the combiner is a single product $\mathbf A_k\mathbf y_k$ at a cost of $\mathcal O(M^2K)$ over all users, or $\mathcal O(MK\log M)$ for the diagonal variants that use the \ac{DFT} matrix to diagonalize the spatial correlation matrices. Computing the \ac{LS} estimate costs $\mathcal{O}(M\tau K)$ over all users, which is the per-realization cost of \ac{MR}-\ac{LS} combining and is added to all other combiners, since each acts on the \ac{LS} estimate. Applying the \ac{MMSE} prefilter $\mathbf R_k\mathbf \Psi_k^{-1}$ onto this costs $\mathcal O(M^2K)$ over all users, which is therefore also the per-realization cost of \ac{MR}-\ac{MMSE} combining. Note that the prefilter $\mathbf R_k\mathbf \Psi_k^{-1}$ itself is formed once in the statistics-based design at a cost of $\mathcal O(M^3K)$, and is also required by \ac{MMSE} combining. On a per-realization level \ac{MMSE} combining inverts a shared $M\times M$ matrix, leading to the total per-realization cost of $\mathcal O(M^3)$, assuming $M$ dominates. Table~\ref{Table:Cost} summarizes these costs. One may note that \ac{DISC}-D and \ac{OBE}-D exhibit a lower per-realization cost than \ac{MR}-\ac{MMSE}, which is due to the fact that they only require point-wise multiplication in the \ac{DFT} domain instead of a full matrix-vector multiplication. 
\begin{table}[t]
\centering
\caption{Computational cost of the considered combiners.}
\label{Table:Cost}
\renewcommand{\arraystretch}{1.3}
\begin{tabular}{l|c|c}
\toprule
Combiner & Statistical filter design & Per-realization \\
\midrule
\ac{DISC}   & $\mathcal{O}(M^3K)$      & $\mathcal{O}(M^2K)$      \\
\ac{DISC}-D & $\mathcal{O}(MK\log M)$  & $\mathcal{O}(MK\log M)$  \\
\ac{OBE}    & $\mathcal{O}(M^3K)$      & $\mathcal{O}(M^2K)$      \\
\ac{OBE}-D  & $\mathcal{O}(MK\log M)$  & $\mathcal{O}(MK\log M)$  \\
\ac{MR}-\ac{MMSE}     & $\mathcal O(M^3K)$                      & $\mathcal{O}(M^2K)$      \\
\ac{MR}-\ac{LS} & -                      & $\mathcal{O}(MK \tau)$      \\
\ac{MMSE}   & $\mathcal O(M^3K)$                     & $\mathcal{O}(M^3)$      \\
\bottomrule
\end{tabular}
\vspace{-1pt}
\end{table}

\begin{figure}[t!]
\centering
\subfloat[Time elapsed for statistical filter design.\label{fig:Cost_design}]{%
  \includegraphics[width=0.9\columnwidth]{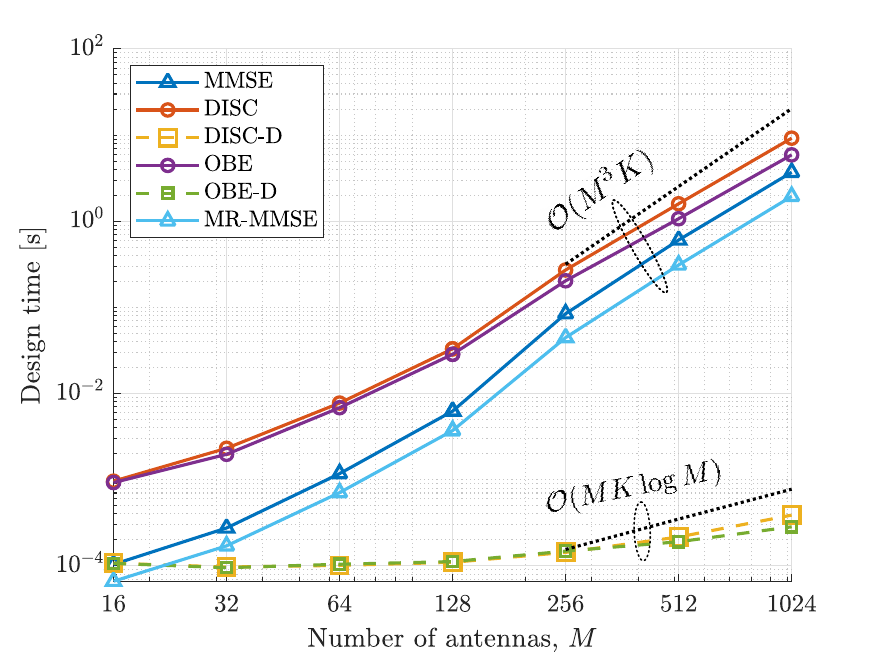}}\\
\subfloat[Time elapsed per-realization.\label{fig:Cost_apply}]{%
  \includegraphics[width=0.9\columnwidth]{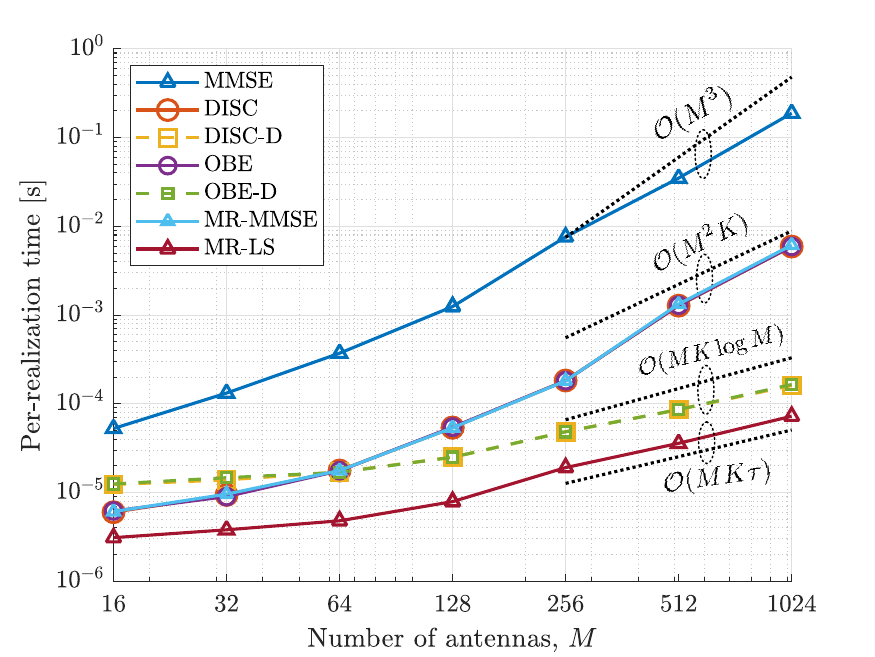}}
\caption{Median time elapsed for each combiner with reference slopes corresponding to the asymptotic complexity in $M$.}
\label{fig:Cost}
\vspace{-4pt}
\end{figure}

To understand how the big-O costs translate into running time, we report the median per-call cost for each combiner's statistical filter design and per-realization cost in Fig.~\ref{fig:Cost_design} and Fig.~\ref{fig:Cost_apply}. Each function is run until either $1000$ calls are made or a $3$-minute budget is reached, and reference slopes are included to show how well the run-time reflects the asymptotic complexity. The exact runtimes will depend on software and implementation details. However, this provides a clear sense of how the combiner costs compare and supports the computational cost analysis in Table~\ref{Table:Cost}, as the reference slopes match the run-time relatively well. The main exception is the design cost of \ac{DISC}-D and \ac{OBE}-D, whose runtime grows more slowly than the reference slope, likely because operations that do not depend on $M$, such as a $K\times K$ inverse, contribute noticeably. In the figure, we observe that the statistical filter design of \ac{DISC} and \ac{OBE} is not significantly more expensive than computing the prefilter for the \ac{MMSE} channel estimate, with the relative difference decreasing as $M$ increases. Note that \ac{MMSE} combining requires computing the uncertainty, $\mathbf Z$, making its design slightly more expensive than \ac{MR}-\ac{MMSE}, even though they both share the $\mathcal{O}(M^3K)$ scaling. On a per-realization level, we can observe that \ac{DISC}, \ac{OBE}, and \ac{MR}-\ac{MMSE} have a similar cost, since they all require an $M\times M$ matrix being multiplied by an $M\times 1$ vector, while \ac{MMSE} combining has a significantly higher cost since it requires an $M\times M$ inverse. It is clear that \ac{DISC}-D and \ac{OBE}-D are considerably less expensive on both the slow-statistics time scale and per realization. The only combiner with a lower cost than these is \ac{MR}-\ac{LS}, which does not apply any interference suppression on pilot-sharing interferers, thus leading to poor performance as shown in Section~\ref{Numerical_results}.
\subsection{Analytical properties and comparisons}
\label{Sec:Analytics_comp}
In this section, we analytically compare \ac{DISC} to \ac{MMSE} and \ac{OBE}, showing when \ac{DISC} is equivalent to the \ac{MMSE} combiner and how it uses more of the available information than \ac{OBE} does. Further, we establish the asymptotic properties of the deterministic \ac{SINR}.

\begin{lemma}\label{MMSE_lemma}
Under the assumption that all users use the same pilot, and with the spatial correlation of every user $k$ given as $\mathbf R_k=\beta_k\mathbf R$, \ac{DISC} is equivalent to the \ac{MMSE} combiner and achieves the same \ac{SINR} for every user.
\end{lemma}
\begin{IEEEproof}
Under these assumptions, the \ac{MMSE} channel estimate becomes $\mathbf{\hat h}_k=\sqrt{\eta_k\tau}\beta_k\mathbf q$, where $\mathbf q\triangleq \mathbf R\mathbf \Psi^{-1}\mathbf y$ is common to all users. The \ac{MMSE} combiner in \eqref{MMSE_comb} can then be written as
\begin{equation}
    \label{MMSE_proof}
p_k\Big(\underbrace{\sum_{i=1}^Kp_i\eta_i\tau \beta_i^2}_{a} \mathbf q\mathbf q^H+\underbrace{\mathbf Z+\sigma^2\mathbf I_M}_{\mathbf G}\Big)^{-1}\sqrt{\eta_k\tau}\beta_k \mathbf q,
\end{equation}
 where $\mathbf G\in \mathbb C^{M\times M}$ is deterministic and positive definite, $a\in \mathbb  R$ is a scalar, and $\mathbf q\in \mathbb C^M$ is a random vector. Applying the Woodbury identity gives
\begin{flalign}
& (a\mathbf q\mathbf q^H+\mathbf G)^{-1}=\mathbf G^{-1}-\mathbf G^{-1}\mathbf q(\frac{1}{a}+\mathbf q^H\mathbf G^{-1}\mathbf q)^{-1}\mathbf q^H\mathbf G^{-1} && \nonumber \\
& =\mathbf G^{-1}-\frac{a\mathbf G^{-1}\mathbf q\mathbf q^H\mathbf G^{-1}}{1+a\rho},
\end{flalign}
with $\rho\triangleq\mathbf q^H\mathbf G^{-1}\mathbf q\in \mathbb R_+$. We now return to \eqref{MMSE_proof} and write $(a\mathbf q\mathbf q^H+\mathbf G)^{-1}\mathbf q$ as
\begin{equation}
\mathbf G^{-1}\mathbf q-\frac{a\mathbf G^{-1}\mathbf q \mathbf q^H\mathbf G^{-1}\mathbf q}{1+a\rho}=\left(\frac{1}{1+a\rho}\right)\mathbf G^{-1}\mathbf q.
\end{equation}
Substituting back and recalling that the scaling of a combiner does not impact the \ac{SINR}, it then follows that
\begin{equation}
\mathbf v_k^{\mathrm{MMSE}}\propto \underbrace{(\mathbf Z+\sigma^2 \mathbf I_M)^{-1}\mathbf R\mathbf \Psi^{-1}}_{\mathbf{\overline A}_k}\mathbf y.
\end{equation}

The \ac{DISC} combiner has the same structure. Introducing $\mathbf q'\triangleq\mathrm{vec}(\mathbf R)$, $\mathbf G'\triangleq\mathbf \Psi^T\otimes(\mathbf Z+\sigma^2\mathbf I_M)$, $a_k'\triangleq\sum_{i\neq k} p_i\eta_i\tau\beta_i^2$, it takes the form
\begin{equation}
\mathbf a_k^\mathrm{DISC}=(\mathbf G'+a_k'\mathbf q'\mathbf q'^H)^{-1}\mathbf q',
\end{equation}
which is identical in structure to \eqref{MMSE_proof}. The scalar $a_k'$ excludes user $k$, unlike in \eqref{MMSE_proof}, but both cancel under scale invariance. The same technique then applies, and we get
\begin{equation}
\mathbf a_k^\mathrm{DISC}\propto \Big(\mathbf \Psi^T\otimes(\mathbf Z+\sigma^2\mathbf I_M)\Big)^{-1}\mathrm{vec}(\mathbf R).
\end{equation}
This equals $\mathrm{vec}((\mathbf Z+\sigma^2\mathbf I_M)^{-1}\mathbf R\mathbf \Psi^{-1})$ by standard Kronecker identities \cite[Lemma~4.3.1]{Horn_Johnson_1991}, and unvectorizing gives
\begin{equation}
\mathbf A_k^{\mathrm{DISC}}=(\mathbf Z+\sigma^2\mathbf I_M)^{-1}\mathbf R\mathbf \Psi^{-1},
\end{equation}
which equals $\mathbf{\overline A}_k$ and closes the proof.
\end{IEEEproof}
The assumptions of Lemma \ref{MMSE_lemma} include the special case of i.i.d. Rayleigh fading, where $\mathbf h_k\sim \mathcal N_\mathbb C(0,\beta_k\mathbf I_M)$. More generally, the condition $\mathbf R_k=\beta_k\mathbf R$ holds whenever the users share the same spatial signature and differ only in the large-scale gain $\beta_k$. This occurs, for example, when users are clustered in a small angular region relative to the \ac{BS}, so that they are seen from a common direction with similar scattering. Grouping statistically equivalent users in this way has also been considered in prior work \cite{6241389}.

Now, let us compare \ac{DISC} and \ac{OBE}. Recalling that \ac{OBE} is obtained by optimizing the \ac{UaTF} bound, the difference between \ac{DISC} and \ac{OBE} stems from \ac{OBE}'s ``\textit{forgetting}'' in that bound. The \ac{DISC} and \ac{OBE} combiners are given by
\begin{flalign}\label{DISC_eq}
& \mathbf a_k^{\mathrm{DISC}}=\Big(\mathbf \Psi_k^T \otimes (\sum_{i\notin\mathcal P_k} p_i\mathbf \Upsilon_i+\sum_{i=1}^K p_i(\mathbf R_i-\mathbf \Upsilon_i)+\sigma^2\mathbf I_M)\ && \nonumber \\
&+\sum_{i\in \mathcal P_k\setminus \{k\}}p_i\eta_i\tau \mathrm{vec}(\mathbf R_i)\mathrm{vec}(\mathbf R_i)^H\Big)^{-1}\mathrm{vec}(\mathbf R_k).
\end{flalign}
\begin{flalign}\label{OBE_eq}
& \mathbf a_k^{\mathrm{OBE}}=\Big(\mathbf \Psi_k^T \otimes (\sum_{i=1}^K p_i\mathbf R_i+\sigma^2\mathbf I_M)\ && \nonumber \\
&+\sum_{i\in \mathcal P_k\setminus \{k\}}p_i\eta_i\tau \mathrm{vec}(\mathbf R_i)\mathrm{vec}(\mathbf R_i)^H\Big)^{-1}\mathrm{vec}(\mathbf R_k).
\end{flalign}
Comparing \eqref{DISC_eq} to \eqref{OBE_eq}, we see that the combiners coincide except for the non-coherent interference, which \ac{OBE} captures as  $\sum_{i=1}^K p_i\mathbf R_i$ while \ac{DISC} obtains $\sum_{i\notin\mathcal P_k} p_i\mathbf \Upsilon_i+\sum_{i=1}^K p_i(\mathbf R_i-\mathbf \Upsilon_i)$. The difference between the combiners is not a result of simply modifying \ac{OBE}, but instead arises from the derivation of the deterministic \ac{SINR}. Through the decomposition in \eqref{LMMSE_y}, \ac{DISC} separates the interference into what $\mathbf y_k$ predicts about $\mathbf y_i$ and what it does not. When users $i$ and $k$ share a pilot, the combiner knows $\mathbf y_i$ fully through $\mathbf y_k$, and the only remaining uncertainty is the unknown part of the channel, captured by $\mathbf R_i-\mathbf \Upsilon_i$. When they do not share a pilot, $\mathbf y_k$ carries no information about $\mathbf y_i$ and $\mathbf \Upsilon_i+(\mathbf R_i-\mathbf \Upsilon_i)$ sums to $\mathbf R_i$, matching \ac{OBE}. By contrast, \ac{OBE} does not account for the correlation between $\mathbf y_k$ and $\mathbf y_i$ since the forgetting in the \ac{UaTF} bound discards it. The sum $\sum_{i=1}^K p_i\mathbf R_i$ also includes $i=k$, so the desired user appears in its own interference term, which introduces self-interference. \ac{DISC} avoids this by accounting for the correlation.

\begin{figure}[t!]
\centerline{\includegraphics[width=0.9\columnwidth]{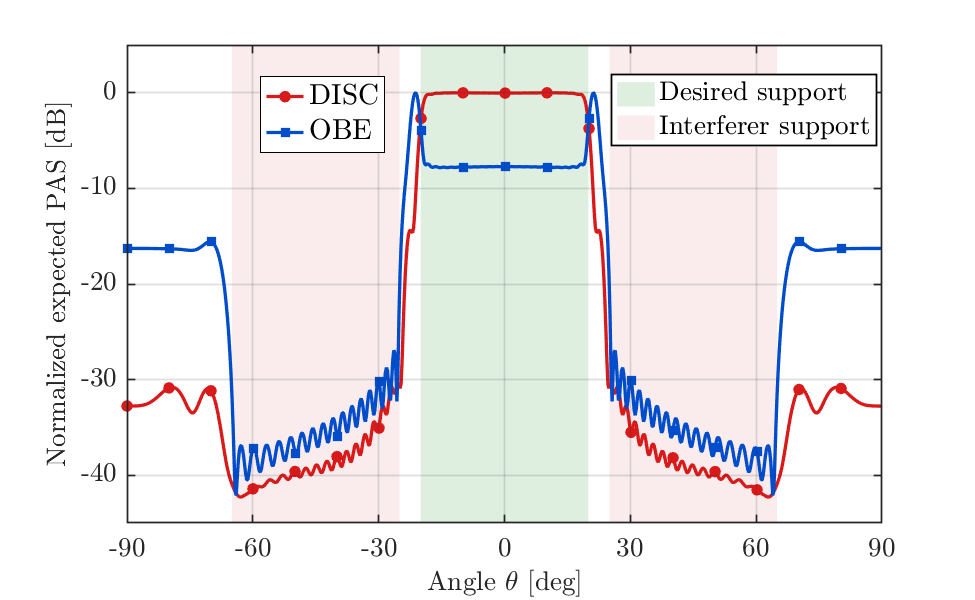}}
\caption{Normalized expected \ac{PAS} of \ac{DISC} and \ac{OBE} under full pilot sharing, with the desired user at nominal angle $0^\circ$ and interferers at $\pm45^\circ$.}
\label{OBE_beams}
\vspace{-10pt}
\end{figure}

To illustrate this, we set the $\mathrm{SNR}=\eta_k\beta_k/\sigma^2$ to $15$ dB and model the scatterers as uniformly distributed over $[\varphi-\Delta, \ \varphi+\Delta]$, with $\Delta=20^\circ$. In Fig. \ref{OBE_beams} we place three users at $\varphi_1=0^\circ$, $\varphi_2=-45^\circ$, and $\varphi_3=45^\circ$ and plot the expected \ac{PAS} of \ac{DISC} and \ac{OBE} for user 1 under full pilot sharing. The expected \ac{PAS} is $\mathbb E[|\mathbf v_k^H\mathbf a(\theta)|^2]$, where $\mathbf a(\theta)$ is the steering vector and the expectation is over the combiner input. The self-interference limits \ac{OBE}'s response in the desired directions, as can be seen in the expected \ac{PAS}. The interference suppression is also stronger with \ac{DISC}.

\begin{remark}\label{Remark_DISC>OBE}
   For every $M$, the optimized \ac{DISC} \ac{SINR} in \eqref{SINR_disc_opt} and the optimized \ac{UaTF} \ac{SINR} of \ac{OBE} differ only in the non-coherent interference term in the denominator. Since $\mathbf \Upsilon_i\succeq 0$ and $\mathbf R_i\succeq 0$ for all $i$, we have
   \begin{equation}
    \sum_{i=1}^Kp_i\mathbf R_i\succeq\sum_{i\notin\mathcal P_k} p_i\mathbf \Upsilon_i+\sum_{i=1}^Kp_i(\mathbf R_i-\mathbf \Upsilon_i),
   \end{equation}
    meaning that \ac{DISC} has the smaller denominator and therefore the larger \ac{SINR}, i.e., $\overline{\mathrm{SINR}}_k^\mathrm{DISC}\geq \mathrm{SINR}_k^\mathrm{UaTF}$. The two quantities are, however, design criteria rather than achieved performance, so this does not necessarily imply that \ac{DISC} outperforms \ac{OBE}. The instantaneous \ac{SINR} that the combiners achieve is evaluated numerically in Section~\ref{Numerical_results}.
\end{remark}

It has been shown that the \ac{SINR} under \ac{MMSE} combining grows unbounded as $M\to \infty$ under suitable conditions on the spatial correlation matrices \cite{Unlimited}. We now show that $\overline{\mathrm{SINR}}^\mathrm{DISC}_k$ shares this property. The condition required is that the pilot-sharing users occupy asymptotically distinct spatial signatures and that the array gathers more energy as $M$ increases, which we state as follows.

\begin{condition}\label{cond:indep}
The spatial correlation matrices of the pilot-sharing users are asymptotically linearly independent, meaning
\begin{equation}\liminf_{M\to \infty} \inf_{\boldsymbol \lambda:\left\|\boldsymbol \lambda\right\|_2=1} \frac{1}{M}\|\sum_{i\in \mathcal P_k} \lambda_i\mathbf R_i\|_F^2>0.\end{equation}
\end{condition}

\begin{condition}\label{cond:grow}
    $\liminf_M \frac{1}{M}\mathrm{tr}(\mathbf R_k)>0.$
\end{condition}

Both conditions are widely used in the massive \ac{MIMO} literature and correspond well to reported real-life measurements \cite{Condition_1,Measurements}. Given these, we formulate the following lemma.

\begin{lemma}\label{lem:unbounded}
Under Condition~\ref{cond:Boundness}, Condition~\ref{cond:indep}, and Condition~\ref{cond:grow}, $\overline{\mathrm{SINR}}^\mathrm{DISC}_k$ grows unbounded as $M\to \infty$.
\end{lemma}
\begin{IEEEproof}
We recall that $\overline{\mathrm{SINR}}^\mathrm{DISC}_k=\eta_k\tau p_k\mathrm{vec}(\mathbf R_k)^H\mathbf M_k^{-1}\mathrm{vec}(\mathbf R_k)$ and introduce $\mathbf \Xi_{-k}\triangleq[\mathrm{vec}(\mathbf R_i)]_{i\in \mathcal P_k\setminus \{k\}}\in\mathbb C^{M^2\times (|\mathcal P_k|-1)}$, which stacks the vectorized correlation matrices of the pilot-sharing interferers. We further define $\mathbf P_{-k}\triangleq \mathrm{diag}(\eta_i\tau p_i)_{i\in {\mathcal P_k\setminus \{k\}}}$, and $\mathbf G_k'\triangleq \mathbf \Psi_k^T\otimes (\sum_{i\notin \mathcal P_k}p_i\mathbf \Upsilon_i+\mathbf Z+\sigma^2\mathbf I_M)$, so that
\begin{equation}
\label{Woodbury_M}
\mathbf M_k = \mathbf G_k' + \mathbf\Xi_{-k}\mathbf P_{-k}\mathbf\Xi_{-k}^H.
\end{equation}
The matrix $\mathbf G_k'$ is positive definite and, by Condition~\ref{cond:Boundness}, spectrally bounded. Applying the Woodbury identity to $\mathbf M_k^{-1}$ and introducing $\mathbf u_k\triangleq\mathbf G_k'^{-1/2}\mathrm{vec} (\mathbf R_k)$ and $\mathbf F_{-k}\triangleq\mathbf G_k'^{-1/2}\mathbf \Xi_{-k}$, we write $\mathrm{vec}(\mathbf R_k)^H\mathbf M_k^{-1}\mathrm{vec}(\mathbf R_k)$ as
\begin{equation}
    \mathbf u_k^H\mathbf u_k-\mathbf u_k^H\mathbf F_{-k}(\mathbf P_{-k}^{-1}+\mathbf F_{-k}^H\mathbf F_{-k})^{-1}\mathbf F_{-k}^H\mathbf u_k.
\end{equation}
Since $\mathbf P_{-k}\succ0$ we have $(\mathbf P_{-k}^{-1}+\mathbf F_{-k}^H\mathbf F_{-k})^{-1}\preceq(\mathbf F_{-k}^H\mathbf F_{-k})^{-1}$. It then follows that
\begin{equation}
\overline{\mathrm{SINR}}^\mathrm{DISC}_k\geq \eta_k\tau p_k\mathbf u_k^H(\mathbf I_{M^2}-\mathbf F_{-k}(\mathbf F_{-k}^H\mathbf F_{-k})^{-1}\mathbf F_{-k}^H)\mathbf u_k.    
\end{equation}
The matrix $\mathbf I_{M^2}-\mathbf F_{-k}(\mathbf F_{-k}^H\mathbf F_{-k})^{-1}\mathbf F_{-k}^H$ is the orthogonal projector onto the complement of $\mathrm{range}(\mathbf F_{-k})$, meaning that the quadratic form is the squared distance from $\mathbf u_k$ to that range, and we can formulate the \ac{SINR} as
\begin{equation}
\overline{\mathrm{SINR}}^\mathrm{DISC}_k\geq \eta_k\tau p_k\min _\mathbf c\|\mathbf u_k-\mathbf F_{-k}\mathbf c\|^2.    
\end{equation}
Using that $\|\mathbf G_k'^{-1/2}\mathbf x \|^2\geq \frac{\|\mathbf x\|^2}{\|\mathbf G_k'\|_2}$, for any vector $\mathbf x$, gives
\begin{equation}
    \overline{\mathrm{SINR}}^\mathrm{DISC}_k\geq \frac{\eta_k\tau p_k}{\|\mathbf G_k'\|_2}\min_{\mathbf c}\big\|\mathrm{vec}(\mathbf R_k)-\mathbf\Xi_{-k}\mathbf c\big\|^2.
\end{equation}
We have now expressed the \ac{SINR} as the squared distance from $\mathrm{vec}(\mathbf R_k)$ to the span of spatial correlation matrices of the pilot-sharing interferers. Since Condition~\ref{cond:grow} ensures that $\|\mathrm{vec}(\mathbf R_k)\|^2$ grows with order of $M$, Condition~\ref{cond:indep} gives that the distance $d_k\triangleq\min_{\mathbf c}\big\|\mathrm{vec}(\mathbf R_k)-\mathbf\Xi_{-k}\mathbf c\big\|^2$ grows with order of $M$. By recalling that the spectral norm of $\mathbf G_k'$ is bounded, we then get
\begin{equation}
    \liminf_M\frac{\overline{\mathrm{SINR}}^\mathrm{DISC}_k}{M}\geq \liminf_M\frac{\eta_k\tau p_k}{\|\mathbf G_k'\|_2}\frac{d_k}{M}>0,
\end{equation}
which closes the proof.
\end{IEEEproof}

Since the deterministic $\overline{\mathrm{SINR}}_k$ was constructed term by term from the instantaneous \ac{SINR}, it is natural to ask whether the two agree asymptotically as $M\to \infty$. We now show that their ratio tends to $1$ almost surely when pilot contamination is present, thereby establishing a meaningful connection between the deterministic \ac{SINR} and the true, instantaneous \ac{SINR}. Together with Lemma~\ref{lem:unbounded}, this also implies that the instantaneous \ac{SINR} under \ac{DISC} grows unbounded in $M$.

\begin{lemma}\label{lem:Ratios}
Let $\mathbf v_k=\mathbf A_k\mathbf y_k$ with $\mathbf A_k$ deterministic and spectrally bounded. Under Conditions~\ref{cond:Boundness} and \ref{cond:grow}, and provided that user $k$ shares its pilot with at least one other user, \begin{equation*}
    \overline{\mathrm{SINR
}}_k(\mathbf A_k)/\mathrm{SINR}_k(\mathbf A_k)\to 1 \quad \text{almost surely as } M\to \infty,
\end{equation*}where $\overline{\mathrm{SINR
}}_k$ refers to \eqref{SINR_disc} while $\mathrm{SINR}_k$ refers to the instantaneous \ac{SINR} in \eqref{SINR_inst}.
\end{lemma}
\begin{IEEEproof}
We express the numerator of \eqref{SINR_inst} as
    \begin{equation}\mathbf N_k\triangleq\frac{1}{M^2}p_k|\mathbf y_k^H\mathbf A_k^H\mathbf {\hat h}_k|^2=\overline n_k+\epsilon_k^{(1)},
    \end{equation} where $\overline n_k\triangleq\frac{1}{M^2} p_k\eta_k\tau|\mathrm{tr}(\mathbf A_k^H\mathbf R_k)|^2$ and $\epsilon_k^{(1)}\triangleq\mathbf N_k-\overline n_k$. Similarly, we express the order-$M$ interference $\mathbf D_k^{(1)}$ as
    \begin{flalign}
       & \mathbf D_k^{(1)}\triangleq\frac{1}{M}\Big(\mathbf y_k^H\mathbf A_k^H\mathbf Z\mathbf A_k\mathbf y_k+\sigma^2\mathbf y_k^H\mathbf A_k^H\mathbf A_k\mathbf y_k+     && \nonumber \\ &    \sum_{i\notin \mathcal P_k}p_i|\mathbf y_k^H\mathbf A_k^H\mathbf {\hat h}_i|^2\Big)=
     \overline d_k^{(1)}+\epsilon^{(2)}_k, 
    \end{flalign}
where 
$\overline d_k^{(1)}\triangleq \frac{1}{M}\Big(\sum_{i\notin \mathcal P_k}p_i\mathrm{tr}(\mathbf A_k\mathbf \Psi_k\mathbf A_k^H\mathbf \Upsilon_i)+\mathrm{tr}(\mathbf A_k^H\mathbf Z\mathbf A_k\mathbf \Psi_k)+\sigma^2\mathrm{tr}(\mathbf A_k^H\mathbf A_k\mathbf \Psi_k)\Big)$ and $\epsilon^{(2)}_k\triangleq\mathbf D_k^{(1)}-\overline d_k^{(1)}.$ Finally, we express the coherent interference $\mathbf D_k^{(2)}$ as 
    \begin{equation}
       \mathbf D_k^{(2)}\triangleq\frac{1}{M^2}\Big(\sum_{i\in \mathcal P_k\setminus  \{k\}} p_i|\mathbf y_k^H\mathbf A_k^H \mathbf {\hat h}_i|^2\Big)= 
     \overline d_k^{(2)}+\epsilon^{(3)}_k, 
    \end{equation}
    with $\overline d_k^{(2)}\triangleq \frac{1}{M^2}\Big(\sum_{i\in \mathcal P_k\setminus \{k\}}p_i\eta_i\tau |\mathrm{tr}(\mathbf A_k^H\mathbf R_i)|^2\Big)$ and $\epsilon^{(3)}_k\triangleq\mathbf D_k^{(2)}-\overline d_k^{(2)}.$ With all parts assembled, we can now express the instantaneous \ac{SINR} as
    \begin{equation}
    \label{SINR_inst_eps}
        \mathrm{SINR}_k=\frac{M^2(\overline n_k+\epsilon_k^{(1)})}{M(\overline d_k^{(1)}+\epsilon_k^{(2)})+M^2(\overline{d}_k^{(2)}+\epsilon_k^{(3)})},
    \end{equation}
while the deterministic \ac{SINR} in \eqref{SINR_disc}, in unvectorized form, is
\begin{equation}
    \overline{\mathrm{SINR}}_k=\frac{M^2\overline n_k}{M\overline d_k^{(1)}+M^2\overline d_k^{(2)}}.
\end{equation}
This yields the ratio
\begin{equation}
\label{Ratio}
     \frac{\overline{\mathrm{SINR
}}_k}{\mathrm{SINR}_k}=\frac{\overline{n}_k}{\overline n_k+\epsilon_k^{(1)}}\frac{\frac{1}{M}(\overline d_k^{(1)}+\epsilon_k^{(2)})+(\overline d_k^{(2)}+\epsilon_k^{(3)})}{\frac{1}{M}\overline d_k^{(1)}+\overline d_k^{(2)}}.
\end{equation}

Since $\mathbf A_k$ and the spatial correlation matrices are spectrally bounded under Condition~\ref{cond:Boundness}, the almost-sure convergences established in Section~\ref{DISC_sec} give that $\epsilon_k^{(1)}$ and $\epsilon_k^{(3)}$ vanish almost surely as $M\to \infty$. The non-coherent interference does not concentrate around its mean, so its contribution to $\epsilon_k^{(2)}$ does not tend to zero. Since the error is of order $M$, it is nonetheless asymptotically dominated by the coherent interference $\overline d_k^{(2)}$, which is non-zero as user $k$ shares its pilot with another user. Both factors in \eqref{Ratio} therefore tend to $1$, as $\overline n_k>0$ under Condition~\ref{cond:grow}, which closes the proof.
\end{IEEEproof}
Since $\mathbf A_k^{\mathrm{DISC}}$ is spectrally bounded, it is a particular instance of the combiner class in Lemma~\ref{lem:Ratios}, as is \ac{DISC}-D, the diagonal variant. The conclusion of the lemma therefore holds for both the proposed combiners. The lemma requires pilot contamination, which is precisely the regime where the \ac{SINR} is traditionally limited. Under orthogonal pilots, the deterministic \ac{SINR} still approximates the instantaneous \ac{SINR} well at practical numbers of antennas, as shown in Section~\ref{Numerical_results}.

At finite $M$, the deterministic \ac{SINR} is not expected to align perfectly with the instantaneous one. To check whether the error terms $\epsilon_k^{(1)}$, $\epsilon_k^{(2)}$, and $\epsilon_k^{(3)}$ carry a systematic bias, we take the expectation of each. We observe that $\epsilon_k^{(1)}$ and $\epsilon_k^{(3)}$ involve fourth-order moments of $\mathbf y_k$, while $\epsilon_k^{(2)}$ has zero mean. Using Isserlis's theorem \cite{isserlis1918formula} for the expectation of fourth-order moments and recalling that $\mathbf y_k$ is a zero-mean circularly symmetric complex Gaussian random vector with covariance $\mathbf \Psi_k$, we get
\begin{equation}
    \mathbb E[\epsilon_k^{(1)}]=\frac{p_k}{M^2}\mathrm{tr}(\mathbf A_k^H\mathbf \Upsilon_k\mathbf A_k\mathbf\Psi_k),
\end{equation}
\begin{equation}
    \mathbb E[\epsilon_k^{(2)}]=0,
\end{equation}
\begin{equation}
    \mathbb E[\epsilon_k^{(3)}]=\frac{1}{M^2}\sum_{i\in \mathcal P_k\setminus \{k\}}p_i\mathrm{tr}(\mathbf A_k^H\mathbf \Upsilon_i\mathbf A_k\mathbf \Psi_k).
\end{equation}
Since $\mathbf \Psi_k$, $\mathbf \Upsilon_k$, and $\mathbf \Upsilon_i$ are Hermitian positive semi-definite, we can express the biases as $\mathrm{tr}(\mathbf \Upsilon_k^{1/2}\mathbf A_k\mathbf \Psi_k^{1/2}(\mathbf \Upsilon_k^{1/2}\mathbf A_k\mathbf \Psi_k^{1/2})^H)$ and $\mathrm{tr}(\mathbf \Upsilon_i^{1/2}\mathbf A_k\mathbf \Psi_k^{1/2}(\mathbf \Upsilon_i^{1/2}\mathbf A_k\mathbf \Psi_k^{1/2})^H)$, respectively, and thus conclude that the biases are non-negative. Furthermore, $\epsilon_k^{(1)}$ enters the numerator while $\epsilon_k^{(3)}$ enters the denominator in \eqref{SINR_inst_eps}, meaning the two positive biases push the deterministic \ac{SINR} in opposite directions. As a result, we cannot say that the deterministic \ac{SINR} systematically over- or underestimates the \ac{SINR}. It should instead be viewed as an approximation. The coherent-interference bias vanishes under orthogonal pilots, so the net bias of the deterministic \ac{SINR} is expected to depend on the degree of pilot sharing. Both biases enter at one order below the deterministic parts $\overline n_k$ and $\overline d_k^{(2)}$, implying that they vanish relative to those parts as $M$ grows and confirming that the biases can be considered a finite-$M$ discrepancy.

In this section, we have shown that \ac{DISC} recovers \ac{MMSE} combining in the common-covariance regime, retains the unbounded \ac{SINR} growth in $M$, avoids the self-interference that constrains \ac{OBE}, and that the deterministic \ac{SINR} asymptotically tracks the instantaneous one under pilot contamination. We now explore these properties numerically.

\section{Numerical results}
\label{Numerical_results}
We structure the numerical results by first showing that the deterministic \ac{SINR} of the \ac{DISC} closely matches the instantaneous \ac{SINR} obtained from Monte Carlo realizations, thereby justifying its use as a design criterion. We then compare the \ac{DISC} combiners to \ac{MMSE} combining, \ac{OBE}, \ac{MR}-\ac{MMSE}, and \ac{MR}-\ac{LS} over \ac{SNR} and pilot length. Lastly, we compare the sum \ac{SE} across different spatial correlation models to show that \ac{DISC}'s advantage over \ac{OBE} does not depend on any particular correlation model.

We consider a single-cell system with $K=5$ users distributed around a \ac{BS} equipped with a uniform linear array with half-wavelength antenna spacing. We define the \ac{SNR} for user $k$ as $p_k\beta_k/\sigma^2$, which represents the received \ac{SNR} at each antenna. The pilot power $\eta_k$ is set equal to the data power $p_k$, allowing a fair comparison of the combiners without introducing power allocation. If not stated otherwise, we assume $M=64$ antennas. The spatial correlation model follows the formulation in \eqref{Spat_cor}, where the probability density function $f(\overline\varphi)$ varies depending on the choice of model. One widely considered model is the one-ring model, in which the density is uniform around a nominal angle, meaning that the angle is distributed as $\mathcal{U}[\varphi-\Delta, \ \varphi+\Delta]$, where $\varphi$ is the nominal angle and $\Delta$ is the half-width of the angular support. Other commonly used models include Gaussian or Laplacian densities centered at the nominal angle. In real-life scenarios, however, the scattering may be spread over several disjoint clusters rather than a single one. Based on the stochastic geometry-based channel model defined by \ac{3GPP} \cite{Correlated_fading,3GPP}, we therefore also consider a model in which the correlation matrix is averaged over $N_C=6$ clusters, each with a nominal angle distributed as $\mathcal U[\varphi-40^\circ,\varphi+40^\circ]$, and multipath components Gaussian distributed around it with a standard deviation of $5^\circ$. Unless stated otherwise, we use this \ac{3GPP}-based model. Throughout, the \ac{SE} of each user is obtained as $\mathbb E[\log_2(1+\mathrm{SINR}_k)]$, where the \ac{SINR} of user $k$ is given by \eqref{SINR_inst}, with the expectation evaluated over $200$ Monte Carlo realizations of the small-scale fading. The sum \ac{SE} is then averaged over $50$ random user placements.

\begin{figure}[t!]
\centerline{\includegraphics[width=1.0\columnwidth]{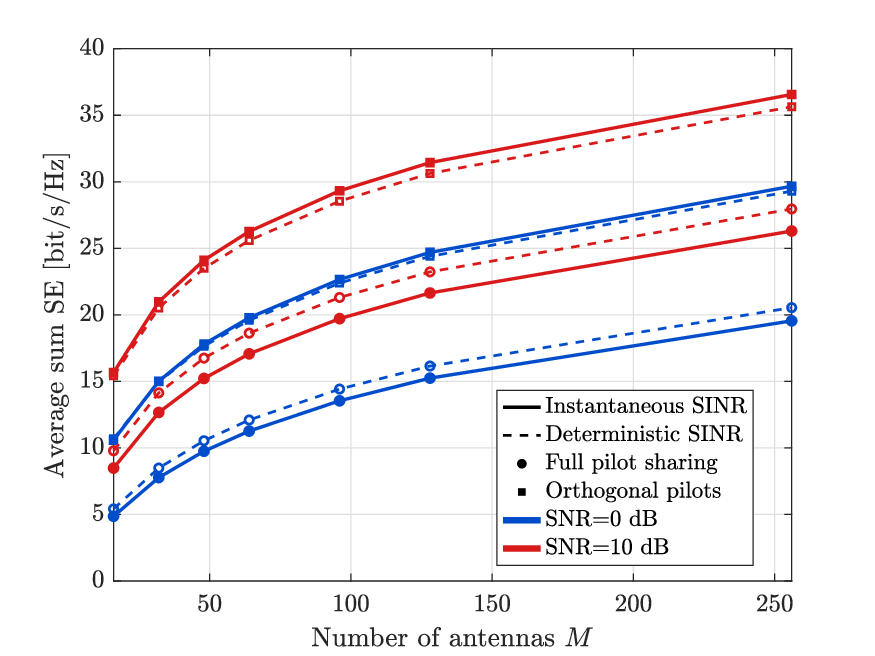}}
\caption{Sum \ac{SE} from the deterministic and instantaneous \ac{SINR} with \ac{DISC} against the number of antennas, at two \ac{SNR} levels under orthogonal pilots and full pilot sharing.}
\label{Validation_figure}
\vspace{-10pt}
\end{figure}

In Fig.~\ref{Validation_figure}, we use \ac{DISC} as the combiner and compare the sum \ac{SE} computed from the deterministic $\overline{\mathrm{SINR}}^{\mathrm{DISC}}_k$ against that computed from the instantaneous \ac{SINR}. Each user's \ac{SNR} is set to either $0$ dB or $10$ dB, to highlight the accuracy at different \ac{SNR} levels, and we show results with both full pilot sharing and mutually orthogonal pilots. The deterministic \ac{SINR} is slightly optimistic under full pilot sharing, while it is slightly pessimistic for orthogonal pilots. This is consistent with the discussion in Section~\ref{Sec:Analytics_comp}, where the finite-$M$ biases for the numerator and coherent interference act in opposite directions on the \ac{SINR}, with the latter appearing only under pilot sharing. Both configurations nonetheless track the instantaneous \ac{SINR} well at a practical number of antennas and across both \ac{SNR} levels. Since the deterministic \ac{SINR} serves as a design criterion rather than an exact performance measure, it is not critical that the two match exactly, only that they follow the same trend. Lemma~\ref{lem:Ratios} establishes this asymptotically under pilot sharing, and Fig.~\ref{Validation_figure} shows that the agreement numerically extends well to orthogonal pilots.

\begin{figure}[t!]
\centerline{\includegraphics[width=0.9\columnwidth]{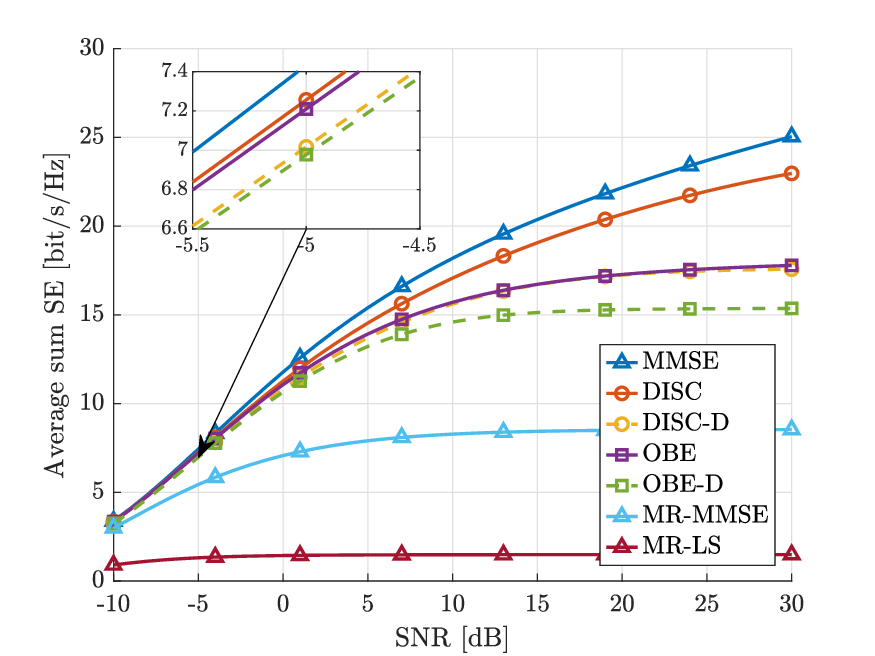}}
\caption{Sum \ac{SE} against \ac{SNR} for the considered combiners under full pilot sharing.}
\label{SNR_fig}
\vspace{-10pt}
\end{figure}

Moving on, in Fig.~\ref{SNR_fig} we compare the combiners over \ac{SNR} under full pilot sharing. Given the strong spatial overlap of the \ac{3GPP} model, \ac{MR}-\ac{MMSE} saturates at high \ac{SNR}, where pilot contamination is the limiting factor. \ac{MR}-\ac{LS} performs even worse, since it combines with the raw \ac{LS} estimate and lacks the prefilter that concentrates the combiner on the desired user's angular support. Turning to the main comparison, \ac{DISC} significantly outperforms \ac{OBE} at high \ac{SNR}, where optimizing against the \ac{UaTF} bound is most costly, since the self-interference it introduces causes the \ac{SINR} to saturate. The gap persists, albeit smaller, at lower \ac{SNR}, indicating that the design of \ac{DISC} is superior to that of \ac{OBE}, consistent with the analytical comparison in Section~\ref{sec:Analytical}. \ac{MMSE} combining serves as the benchmark, and \ac{DISC} achieves a comparable sum \ac{SE} at low \ac{SNR} while, unlike the other combiners, avoiding noticeable saturation at high \ac{SNR}.

\begin{figure}[t!]
\centerline{\includegraphics[width=0.9\columnwidth]{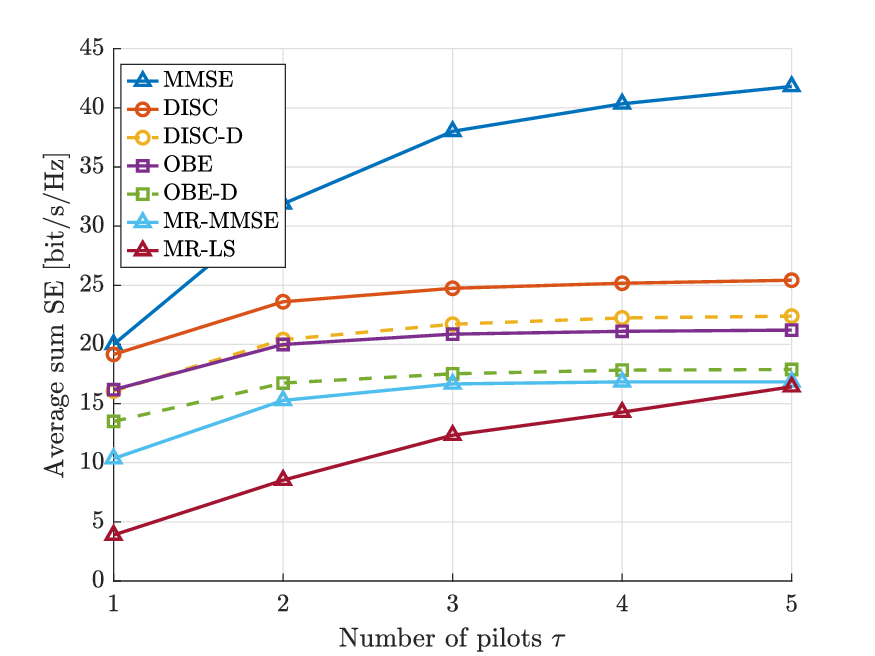}}
\caption{Sum \ac{SE} against the number of pilots $\tau$, using the pilot allocation that maximizes each combiner's deterministic sum \ac{SE}.}
\label{tau_fig}
\vspace{-10pt}
\end{figure}

Fig.~\ref{SNR_fig} showed the performance of \ac{DISC} under full pilot sharing. In Fig.~\ref{tau_fig}, we compare \ac{DISC} to the other combiners over the number of pilots $\tau$, ranging from $\tau=1$, where all users share a single pilot, to $\tau=5$, where the pilots are mutually orthogonal. For each combiner and each $\tau$, we use the pilot allocation that maximizes its corresponding deterministic sum \ac{SE}, found through exhaustive search, while \ac{MMSE}, \ac{MR}-\ac{MMSE}, and \ac{MR}-\ac{LS} combining use the allocation of \ac{DISC} since they do not have a corresponding deterministic sum \ac{SE}. Each user's \ac{SNR} is uniformly drawn from $-10$ dB to $30$ dB. We set the prelog factor to one, so that the comparison reflects the achieved \ac{SINR} rather than the pilot overhead. Firstly, we can see that \ac{MR}-\ac{LS} approaches \ac{MR}-\ac{MMSE} as the pilots become orthogonal, where the downside of using the raw \ac{LS} estimate shrinks. \ac{MMSE} combining increases substantially as the number of pilots increases, since it can instantaneously suppress interference effectively once users no longer share pilots. Notably, \ac{MMSE} combining performs well under the allocation optimized for \ac{DISC}, suggesting that the deterministic \ac{SINR} may serve as a low-complexity criterion for pilot allocation more broadly. \ac{DISC} and \ac{DISC}-D outperform \ac{OBE} and \ac{OBE}-D, respectively, across pilot configurations, and remain effective when few pilots are used. Interestingly, for some user placements, fewer pilots yielded a larger sum \ac{SE} than more pilots for \ac{DISC} and \ac{DISC}-D, even with the prelog factor fixed to one. A related effect was observed previously for \ac{MR}-\ac{MMSE} combining in \cite{wennemo2026exploitingpilotcontaminationimprove}, and it appears to partially carry over to \ac{DISC}.

\begin{figure}[t!]
\centerline{\includegraphics[width=0.9\columnwidth]{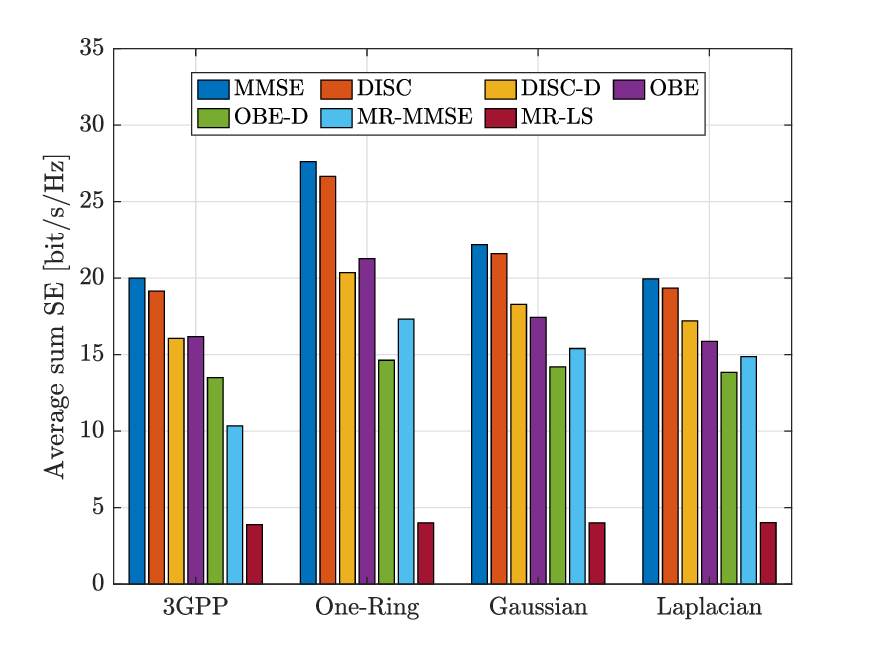}}
\caption{Sum \ac{SE} across different spatial correlation models under full pilot sharing.}
\label{Spat_fig}
\vspace{-10pt}
\end{figure}

In the previous results shown in this section, we have assumed the \ac{3GPP} spatial correlation model to examine how \ac{DISC} performs in scenarios with multiple disjoint clusters, thereby modeling real-world conditions. In Fig.~\ref{Spat_fig}, we highlight how the sum \ac{SE} differs for the combiners across the \ac{3GPP} model, the one-ring model, a Gaussian angle distribution, and a Laplacian angle distribution under full pilot sharing. We assume a standard deviation of $10^\circ$ for all single-cluster models with each user's \ac{SNR} drawn uniformly between $-10$ dB and $30$ dB. Across all four models, \ac{DISC} performs close to \ac{MMSE} and outperforms \ac{OBE}, confirming that its advantage does not depend on a particular correlation structure. The gap between \ac{DISC} and \ac{OBE} is largest for the one-ring model, where the concentrated angular support yields the least channel hardening, implying that the \ac{UaTF} bound is loosest here. The concentrated support also yields the highest overall rates because users occupy distinct supports, reducing spatial overlap and making interference easier to suppress. For the Gaussian and Laplacian models, greater spread in the angular supports makes the \ac{DFT}-based diagonalization in \ac{DISC}-D more accurate, resulting in \ac{DISC}-D even outperforming \ac{OBE}.

\section{Conclusion}
In this work, we proposed a novel statistics-based receive combiner for massive \ac{MIMO} systems, referred to as \ac{DISC}, that maximizes a deterministic approximation of the instantaneous \ac{SINR}, obtained by replacing the random terms with deterministic equivalents. We established that \ac{DISC} is equivalent to \ac{MMSE} combining under certain conditions and that its \ac{SINR} grows unbounded in the number of antennas under standard linear-independence conditions. We then analytically characterized how \ac{DISC} exploits available information that is purposely discarded when designing based on the \ac{UaTF} bound. In terms of complexity, \ac{DISC} matches \ac{OBE}, the statistics-based combiner designed against the \ac{UaTF} bound, while avoiding the per-realization inverse required by \ac{MMSE} combining. We further proposed a diagonal low-complexity variant, referred to as \ac{DISC}-D. Finally, we numerically demonstrated that \ac{DISC} outperforms \ac{OBE} across \ac{SNR}, pilot length, and spatial correlation model, and that it performs close to \ac{MMSE} combining under full pilot sharing, while \ac{DISC}-D retains much of this performance at reduced complexity. 

Several directions remain for future work, including scalable implementation across different cell structures such as cell-free \ac{MIMO}, pilot allocation algorithms, and performance under imperfect covariance knowledge. \ac{DISC} could further be extended to the downlink, where power allocation would also be an essential part of the design. More broadly, this work provides a principled alternative to \ac{UaTF}-based combiner design, replacing a heuristic deterministic substitution with a combiner derived directly from the instantaneous \ac{SINR}. Beyond the combiner itself, these results show that users can share pilots even in dense networks while maintaining high performance, provided the pilot-sharing users have sufficiently different spatial correlation matrices. This becomes increasingly relevant in future \ac{MIMO} systems, where a large number of antennas allows many users to be served simultaneously.

\section*{Appendix A - Reducing computational cost}
\label{Appendix:Computation}
To reduce the computational cost of \ac{DISC}, we introduce $\mathbf \Xi_{-k}\triangleq[\mathrm{vec}(\mathbf R_i)]_{i\in \mathcal P_k\setminus \{k\}}$, $\mathbf P_{-k}\triangleq \mathrm{diag}(\eta_i\tau p_i)_{i\in {\mathcal P_k\setminus \{k\}}}$, and $\mathbf G_k'\triangleq \mathbf \Psi_k^T\otimes (\sum_{i\notin \mathcal P_k}p_i\mathbf \Upsilon_i+\mathbf Z+\sigma^2\mathbf I_M)$, so that
\begin{equation}
\mathbf M_k = \mathbf G_k' + \mathbf\Xi_{-k}\mathbf P_{-k}\mathbf\Xi_{-k}^H.
\end{equation}
The naive computation of $\mathbf a_k^\mathrm{DISC}\propto \mathbf M_k^{-1}\mathrm{vec}(\mathbf R_k)$ requires inverting $\mathbf M_k\in \mathbb C^{M^2\times M^2}$ at a cost of $\mathcal O(M^6)$. The Woodbury identity expresses $\mathbf M_k^{-1}$ as the difference of $\mathbf G_k'^{-1}$ and the term $\mathbf G_k'^{-1}\mathbf \Xi_{-k}(\mathbf P^{-1}_{-k}+\mathbf \Xi_{-k}^H\mathbf G_k'^{-1}\mathbf \Xi_{-k})^{-1}\mathbf \Xi_{-k}^H\mathbf G_k'^{-1}$, and we treat each in turn.

For $\mathbf G_k'^{-1}\mathrm{vec}(\mathbf R_k)$, the Kronecker structure gives
\begin{flalign}
\label{Woodbury_solve}
        &(\mathbf \Psi_k^T\otimes (\sum_{i\notin \mathcal P_k}p_i\mathbf \Upsilon_i+\mathbf Z+\sigma^2\mathbf I_M))^{-1}\mathrm{vec}(\mathbf R_k)=
        && \nonumber \\
        &\mathrm{vec}((\sum_{i\notin \mathcal P_k}p_i\mathbf \Upsilon_i+\mathbf Z+\sigma^2\mathbf I_M)^{-1}\mathbf R_k\mathbf \Psi_k^{-1}).
\end{flalign}
The inverse of the $M^2\times M^2$ matrix is never formed, and two $M\times M$ matrix inverses instead suffice at a cost of $\mathcal O(M^3)$.

Moving on to the second term, we notice that the only inverse other than $\mathbf G_k'^{-1}$ is that of the $r\times r$ matrix $\mathbf P_{-k}^{-1}+\mathbf \Xi_{-k}^H\mathbf G_k'^{-1}\mathbf \Xi_{-k}$, where $r=|\mathcal P_k|-1$ is the number of pilot-sharing users excluding user $k$. Applying the identity in \eqref{Woodbury_solve} to each column, $\mathbf G_k'^{-1}\mathbf \Xi_{-k}$ costs $\mathcal O(M^3)$ per interferer and $\mathcal O(M^3r)$ in total. Forming $\mathbf \Xi_{-k}^H\mathbf G_k'^{-1}\mathbf \Xi_{-k}$ then costs $\mathcal O(M^2r^2)$, and the $r\times r$ inverse costs $\mathcal O(r^3)$. Reusing $\mathbf G_k'^{-1}\mathbf \Xi_{-k}$ and this $r\times r$ inverse, we can compute the full expression
\begin{equation}
    \mathbf G_k'^{-1}\mathbf \Xi_{-k}(\mathbf P^{-1}_{-k}+\mathbf \Xi_{-k}^H\mathbf G_k'^{-1}\mathbf \Xi_{-k})^{-1}\mathbf \Xi_{-k}^H\mathbf G_k'^{-1}\mathrm{vec}(\mathbf R_k),
\end{equation}
for an additional cost of $\mathcal O(M^2r)$ and $\mathcal O(r^2)$. Since $r<K<M$, the dominating cost of computing $\mathbf a_k^{\mathrm{DISC}}$ is $\mathcal{O}(M^3r)$. Within a pilot group, the matrix $\mathbf G_k'$ is shared, so the products $\mathbf G_k'^{-1}\mathrm{vec}(\mathbf R_i)$ for $i\in \mathcal P_k$ are computed once and reused, yielding a total design cost of $\mathcal O(M^3K)$ for all users.

An analogous reduction applies to \ac{DISC}-D. In \ac{DISC}-D, the matrix $\mathbf M_k^D$ in \eqref{SINR_disc_d} is already $M\times M$. Diagonalizing the spatial correlation matrices via the \ac{DFT} in advance, as in Section~\ref{sec:DISC_D}, makes $\mathbf M_k^D$ diagonal plus low-rank, so the combiner for all users is obtained via the Woodbury identity at a total cost of just $\mathcal{O}(MK)$, assuming $M$ dominates. The cost is instead dominated by the \ac{DFT} transforms of the spatial correlation matrices at $\mathcal{O}(MK\log M)$ for all users. The same cost applies when the combiner is applied to an instantaneous realization, since the \ac{DFT} transform is then also performed on the received signal. 

\bibliographystyle{IEEEtran}
\bibliography{IEEEabrv,references}

@STRING{IEEE_J_SPL        = "{IEEE} Signal Process. Lett."}

@STRING{IEEE_J_SP         = "{IEEE} Trans. Signal Process."}

@STRING{IEEE_J_COML       = "{IEEE} Commun. Lett."}

@STRING{IEEE_J_COM        = "{IEEE} Trans. Commun."}

@STRING{IEEE_J_WCOM       = "{IEEE} Trans. Wireless Commun."}

@ARTICLE{OBE,
 author={Neumann, David and Wiese, Thomas and Joham, Michael and Utschick, Wolfgang},
  journal=IEEE_J_SP,
  title={A Bilinear Equalizer for Massive {MIMO} Systems},
  year={2018},
  volume={66},
  number={14},
  pages={3740-3751},
  doi={10.1109/TSP.2018.2838577}
}

@ARTICLE{GMR,
  author={Polegre, Alberto {\'A}lvarez and Sanguinetti, Luca and Armada, Ana Garc{\'i}a},
  journal=IEEE_J_COML, 
  title={Pilot Decontamination Processing in Cell-Free Massive {MIMO}}, 
  year={2021},
  volume={25},
  number={12},
  pages={3990-3994},
  doi={10.1109/LCOMM.2021.3118890}}

@INPROCEEDINGS{wang2026statisticsapproximationenableddistributedbeamforming,
   author={Wang, Zhe and Björnson, Emil and Zhang, Jiayi and Zhang, Peng and Petrov, Vitaly and Ai, Bo},
  booktitle={Proc. IEEE ICC}, 
  title={Statistics Approximation-Enabled Distributed Beamforming for Cell-Free Massive {MIMO}}, 
  year={2026},
  volume={},
  number={},
  doi={10.1109/ICC59461.2026.11588270}}

@unpublished{DISC_D,
  author = {Emanuel Wennemo and Ki Won Sung and Emil Björnson},
  title  = {An Enhanced Low-Complexity Combiner for Massive {MIMO} using Deterministic Equivalents},
  note   = {Submitted to IEEE ICC 2027},
  year   = {2026}
}

@ARTICLE{Cell_Free_OBE,
  author={Wang, Zhe and Zhang, Jiayi and Björnson, Emil and Niyato, Dusit and Ai, Bo},
  journal=IEEE_J_SP, 
  title={Optimal Bilinear Equalizer for Cell-Free Massive {MIMO} Systems Over Correlated Rician Channels}, 
  year={2025},
  volume={73},
  number={},
  pages={1501-1517},
  doi={10.1109/TSP.2025.3547380}}

@misc{syed2024designmultiuserrisaidedstatistical,
      title={Design of a Multi-User {RIS}-Aided System with Statistical Channel Knowledge}, 
      author={Sadaf Syed and Dominik Semmler and Donia Ben Amor and Michael Joham and Wolfgang Utschick},
      year={2024},
      eprint={2405.08585},
      archivePrefix={arXiv},
      primaryClass={eess.SP},
      url={https://arxiv.org/abs/2405.08585}, 
}

@phdthesis{Correlated_fading,
  author={{\"O}zdogan, {\"O}zgecan},
  title={Signal Processing Aspects of Massive {MIMO} and {IRS}-Aided Communications},
  year={2022},
  school = {Linköping University}
}

@article{massivemimobook,
  author = {Emil Bj\"{o}rnson and Jakob Hoydis and Luca Sanguinetti},
  title = {Massive {MIMO} Networks: {Spectral}, Energy, and Hardware Efficiency},
  journal = {Foundations and Trends in Signal Processing},
  year = {2017},
  volume = {11},
  number = {3-4},
  pages = {154-655},
  doi = {10.1561/2000000093}
}

@ARTICLE{Unlimited,
  author={Bj\"{o}rnson, Emil and Hoydis, Jakob and Sanguinetti, Luca},
  journal=IEEE_J_WCOM, 
  title={Massive {MIMO} Has Unlimited Capacity}, 
  year={2018},
  volume={17},
  number={1},
  pages={574-590},
  doi={10.1109/TWC.2017.2768423}}

@ARTICLE{Marzetta_intro,
  author={Marzetta, Thomas L.},
  journal=IEEE_J_WCOM, 
  title={Noncooperative Cellular Wireless with Unlimited Numbers of Base Station Antennas}, 
  year={2010},
  volume={9},
  number={11},
  pages={3590-3600},
  doi={10.1109/TWC.2010.092810.091092}}

@article{Demir_2021,
  author = {Demir, {\"O}zlem Tu{\u{g}}fe and Bj\"{o}rnson, Emil and Sanguinetti, Luca},
  title = {Foundations of User-Centric Cell-Free Massive {MIMO}},
  journal = {Foundations and Trends in Signal Processing},
  year = {2021},
  volume = {14},
  number = {3-4},
  pages = {162--472},
  doi = {10.1561/2000000109}
}

@ARTICLE{Marzetta_pilotcon,
  author={Jose, Jubin and Ashikhmin, Alexei and Marzetta, Thomas L. and Vishwanath, Sriram},
  journal=IEEE_J_WCOM, 
  title={Pilot Contamination and Precoding in Multi-Cell {TDD} Systems}, 
  year={2011},
  volume={10},
  number={8},
  pages={2640-2651},
  doi={10.1109/TWC.2011.060711.101155}}

@ARTICLE{Scalable,
  author={Bj\"{o}rnson, Emil and Sanguinetti, Luca},
  journal=IEEE_J_COM, 
  title={Scalable Cell-Free Massive {MIMO} Systems}, 
  year={2020},
  volume={68},
  number={7},
  pages={4247-4261},
  doi={10.1109/TCOMM.2020.2987311}}

@ARTICLE{Towards,
  author={Sanguinetti, Luca and Bj\"{o}rnson, Emil and Hoydis, Jakob},
  journal=IEEE_J_COM, 
  title={Toward Massive {MIMO} 2.0: Understanding Spatial Correlation, Interference Suppression, and Pilot Contamination}, 
  year={2020},
  volume={68},
  number={1},
  pages={232-257},
  doi={10.1109/TCOMM.2019.2945792}
  }

@article{FR3,
  author = {Bj\"{o}rnson, Emil and Kara, Ferdi and Kolomvakis, Nikolaos and Kosasih, Alva and Ramezani, Parisa and Salman, Murat Babek},
  title = {Enabling 6{G} Performance in the Upper Mid-Band by Transitioning From Massive to Gigantic {MIMO}},
  journal = {IEEE Open Journal of the Communications Society},
  year = {2025},
  volume = {6},
  pages = {5450--5463},
  doi = {10.1109/OJCOMS.2025.3576931}
}

@ARTICLE{caire2017ergodicratelowerbounds,
  author={Caire, Giuseppe},
  journal=IEEE_J_WCOM, 
  title={On the Ergodic Rate Lower Bounds With Applications to Massive {MIMO}}, 
  year={2018},
  volume={17},
  number={5},
  pages={3258-3268},
  doi={10.1109/TWC.2018.2808522}}

@ARTICLE{Covariance_estimate_neumann,
  author={Neumann, David and Joham, Michael and Utschick, Wolfgang},
  journal=IEEE_J_SPL, 
  title={Covariance Matrix Estimation in Massive {MIMO}}, 
  year={2018},
  volume={25},
  number={6},
  pages={863-867},
  doi={10.1109/LSP.2018.2827323}}

@ARTICLE{Covariance_estimate_Upadhya,
  author={Upadhya, Karthik and Vorobyov, Sergiy A.},
  journal=IEEE_J_SPL, 
  title={Covariance Matrix Estimation for Massive {MIMO}}, 
  year={2018},
  volume={25},
  number={4},
  pages={546-550},
  doi={10.1109/LSP.2018.2805725}}

@ARTICLE{Condition_1,
  author={Hoydis, Jakob and ten Brink, Stephan and Debbah, Merouane},
  journal={IEEE Journal on Selected Areas in Communications}, 
  title={Massive {MIMO} in the {UL}/{DL} of Cellular Networks: How Many Antennas Do We Need?}, 
  year={2013},
  volume={31},
  number={2},
  pages={160-171},
  doi={10.1109/JSAC.2013.130205}}

@book{Horn_Johnson_2012, place={Cambridge}, edition={2}, title={Matrix Analysis}, publisher={Cambridge University Press}, author={Horn, Roger A. and Johnson, Charles R.}, year={2012}}

@book{gallager2013stochastic,
  title={Stochastic Processes: Theory for Applications},
  author={Gallager, R.G.},
  isbn={9781107435315},
  lccn={81001607},
  year={2013},
  publisher={Cambridge University Press}
}

@article{Szego,
author = {Gray, Robert},
year = {2001},
month = {10},
pages = {},
title = {Toeplitz and Circulant Matrices: A Review},
volume = {2},
journal = {Foundations and Trends in Communications and Information Theory},
doi = {10.1561/0100000006}
}

@book{Horn_Johnson_1991, place={Cambridge}, title={Topics in Matrix Analysis}, publisher={Cambridge University Press}, author={Horn, Roger A. and Johnson, Charles R.}, year={1991}}

@ARTICLE{Measurements,
  author={Gao, Xiang and Edfors, Ove and Tufvesson, Fredrik and Larsson, Erik G.},
  journal=IEEE_J_COM, 
  title={Massive {MIMO} in Real Propagation Environments: Do All Antennas Contribute Equally?}, 
  year={2015},
  volume={63},
  number={11},
  pages={3917-3928},
  doi={10.1109/TCOMM.2015.2462350}}

@ARTICLE{40,
  author={Adhikary, Ansuman and Ashikhmin, Alexei and Marzetta, Thomas L.},
  journal=IEEE_J_COM, 
  title={Uplink Interference Reduction in Large-Scale Antenna Systems}, 
  year={2017},
  volume={65},
  number={5},
  pages={2194-2206},
  doi={10.1109/TCOMM.2017.2662023}}

@INPROCEEDINGS{25000,
  author={Björnson, Emil and Sanguinetti, Luca and Debbah, Merouane},
  booktitle={2016 50th Asilomar Conference on Signals, Systems and Computers}, 
  title={Massive {MIMO} with imperfect channel covariance information}, 
  year={2016},
  volume={},
  number={},
  pages={974-978},
  doi={10.1109/ACSSC.2016.7869195}}

@techreport{3GPP,
  author      = {{3rd Generation Partnership Project (3GPP)}},
  title       = {{Technical Specification Group Radio
Access Network; Spatial channel model for Multiple Input Multiple Output ({MIMO}) simulations}},
  institution = {{3GPP}},
  number      = {TS 25.996},
  version     = {14.0.0},
  year        = {2017},
}

@INPROCEEDINGS{wennemo2026exploitingpilotcontaminationimprove,
      author={Emanuel Wennemo and Ki Won Sung and Emil Björnson},
      title={Exploiting Pilot Contamination to Improve {UL} Sum Rate under Maximum Ratio Combining}, 
  booktitle={Proc. IEEE SPAWC}, 
  year={2026},
  volume={},
  number={},
  doi={}
  }

@article{isserlis1918formula,
  title={On a formula for the product-moment coefficient of any order of a normal frequency distribution in any number of variables},
  author={Isserlis, Leon},
  journal={Biometrika},
  volume={12},
  number={1/2},
  pages={134--139},
  year={1918},
  publisher={JSTOR}
}

@ARTICLE{6241389,
  author={Huh, Hoon and Caire, Giuseppe and Papadopoulos, Haralabos C. and Ramprashad, Sean A.},
  journal=IEEE_J_WCOM, 
  title={Achieving "Massive {MIMO}" Spectral Efficiency with a Not-so-Large Number of Antennas}, 
  year={2012},
  volume={11},
  number={9},
  pages={3226-3239},
  doi={10.1109/TWC.2012.070912.111383}}

\newpage

\vfill

\end{document}